\documentclass[epj]{svjour}
\pdfoutput=1 

\usepackage{amsmath}
\usepackage{amssymb} 
\usepackage{graphicx}

\begin{document}

\title{Google matrix analysis of the multiproduct world trade network} 

\author{L.Ermann$^1$ \and D.L.Shepelyansky$^{2}$}

\institute{
Departamento de F\'isica Te\'orica, GIyA, CNEA, Av. Libertador 8250, 
(C1429BNP) Buenos Aires, Argentina.
\and
Laboratoire de Physique Th\'eorique du CNRS, IRSAMC, 
Universit\'e de Toulouse, UPS, F-31062 Toulouse, France
}

\titlerunning{Google matrix analysis of the multiproduct world trade network}
\authorrunning{L.Ermann and D.L.Shepelyansky}

\abstract{
Using the United Nations COMTRADE database \cite{comtrade} 
we construct the Google matrix $G$ of 
multiproduct world trade between the UN countries 
and analyze the properties of trade flows on this network
for years 1962 - 2010. This construction, based on Markov chains,
treats all countries on equal democratic grounds independently of their 
richness and at the same time it considers the contributions of trade products
proportionally to their trade volume. We consider the trade with 61 products 
for up to 227 countries. The obtained results show that the trade 
contribution of products is asymmetric: some of them are export oriented while 
others are import oriented even if the ranking by their trade volume is 
symmetric in respect to export and import after averaging 
over all world countries.
The construction of the Google matrix allows to 
investigate the sensitivity of 
trade balance in respect to price variations of  products, 
e.g. petroleum and gas,
taking into account the world connectivity of trade links.
The trade balance based on PageRank and CheiRank probabilities 
highlights the leading role of China and other BRICS countries
in the world trade in recent years. We also show that the eigenstates of $G$
with large eigenvalues
select specific trade communities.
}



\date{Dated:  January 14, 2015}

\maketitle

\section{Introduction}

\begin{figure}[!ht]
\begin{center}
\includegraphics[width=0.48\textwidth]{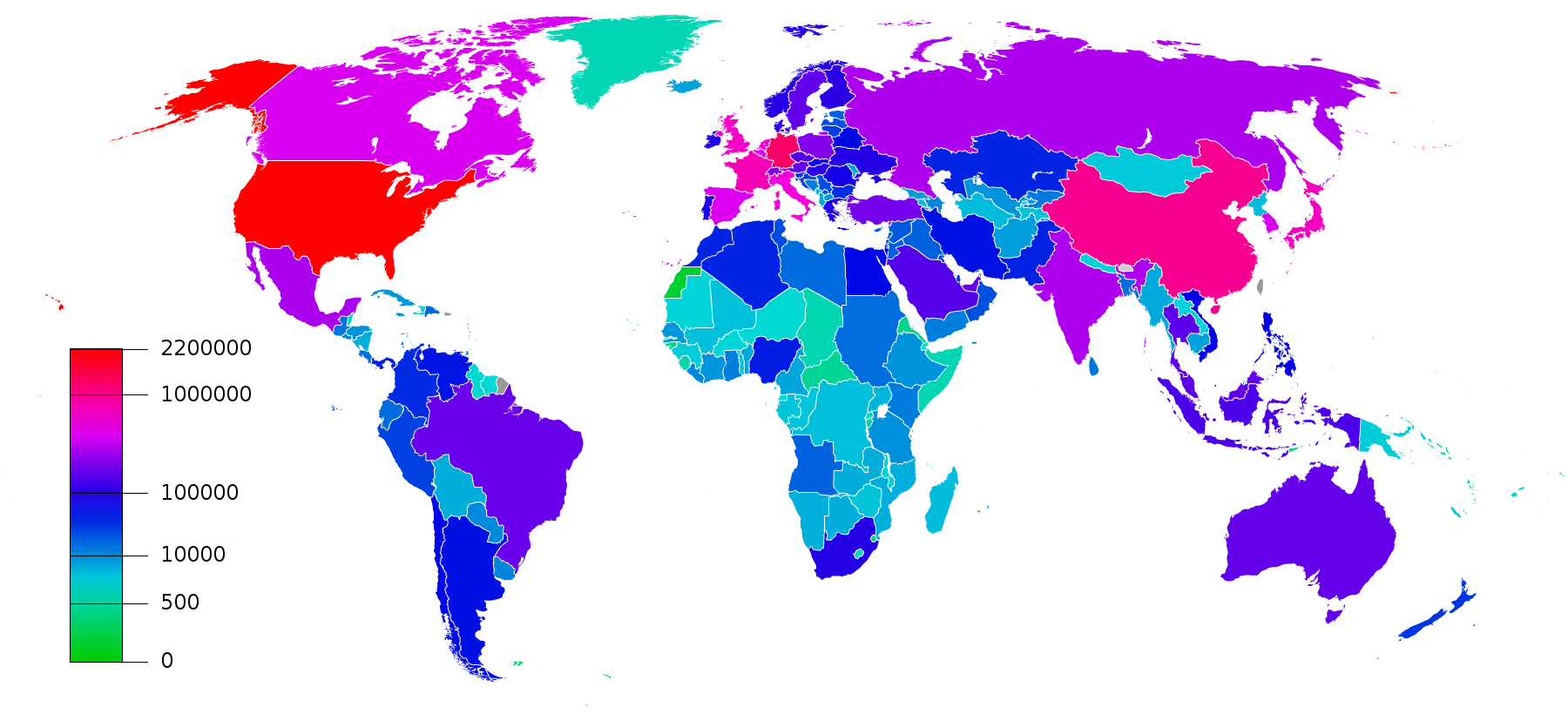}\\
\includegraphics[width=0.48\textwidth]{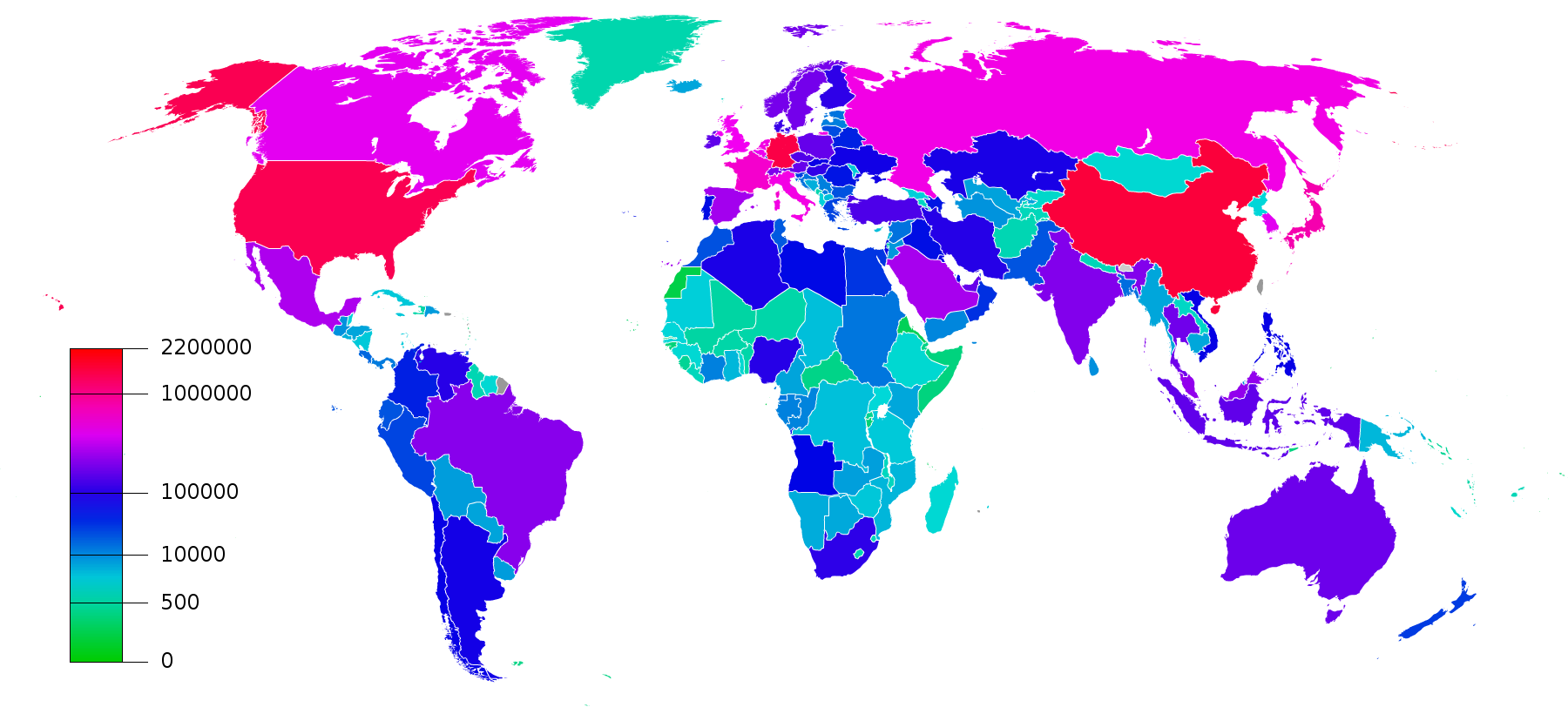}
\end{center}
\vglue -0.1cm
\caption{
World map of countries with color showing country import (top panel)
and export (bottom panel) trade volume expressed in millions of USD
given by numbers of the color bars. The data are shown for year 2008
with $N_c=227$ countries for trade in all $N_p=61$ products (from UN COMTRADE
\cite{comtrade}). Names of countries can be find at
\cite{worldmap}.
}
\label{fig1}
\end{figure}

According to the data of UN COMTRADE \cite{comtrade}
and the international trade statistics 2014 of the World Trade Organization (WTO)
\cite{wto2014}
the international world trade between world countries 
demonstrates a spectacular growth with 
an increasing trade volume and
number of trade products.
It is well clear that the world trade plays the fundamental role in the 
development of world economy \cite{krugman2011}. 
According to the WTO Chief Statistician Hubert Escaith
``In recent years we have seen growing demand for data on the world economy
and on international trade in particular. This demand has grown in particular 
since the 2008-09 crisis, whose depth and breadth 
surprised many experts'' \cite{wto2014}.
In global the data of the world trade exchange
can be viewed as a large multi-functional
directed World Trade Network (WTN) which provides important information 
about multiproduct commercial flows between countries
for a given year. At present the 
COMTRADE database contains data for $N_c=227$ UN countries
with up to $N_p \approx 10^4$ trade products.
Thus the whole matrix of these directed trade flows
has a rather large size $N=N_p N_c \sim 10^6$.
A usual approach is to consider the export and import volumes,
expressed in US dollars (USD). An example of the world map
of countries characterized by their import and export 
trade volume for year
2008 is shown in Fig.~\ref{fig1}.
However, such an approach gives
only an approximate description of trade
where hidden links and interactions between
certain countries and products are not taken into account
since only a country global import or export
are considered.
Thus the  statistical analysis of these 
multiproduct trade data
requires a utilization of more 
advanced mathematical and numerical methods.

In fact, in the last decade, modern societies developed 
enormous communication and social networks including the World Wide Web (WWW),
Wikipedia, Twitter etc. (see e.g. \cite{dorogovtsev}). 
A necessity of information retrieval
from such networks led to a development of efficient algorithms 
for information analysis on such networks appeared in computer science. 
One of the most spectacular tools is the PageRank algorithm developed by
Brin and Page in 1998 \cite{brin}, which became a mathematical foundation
of the Google search engine (see e.g. \cite{meyer}). This algorithm 
is based on  the concept of 
Markov chains and a construction of the Google matrix $G$ of 
Markov transitions between network nodes. The right eigenvector 
of this matrix $G$, known as PageRank vector,
allows to rank all nodes according to their importance 
and influence on the network. The studies of various directed
networks showed that it is useful to analyze also the matrix $G^*$
constructed for the same network but with an inverted direction of links
\cite{linux,wikizzs}. The PageRank vector of $G^*$ is known as the CheiRank vector.
The spectral properties of Goggle matrix for various networks are
described in \cite{arxivrmp}.

The approach of Google matrix to the analysis of 
WTN was started in \cite{wtngoogle}.
The striking feature of this approach is that it treats all UN countries
on equal  democratic grounds, independently of richness of a given country,
in agreement with the principles of UN where all countries are equal.
This property of $G$ matrix is based on the property of Markov chains
where the total probability is conserved to be unity since the sum
of elements for each column of $G$ is equal to unity.
Even if in this approach all countries are treated on equal grounds still
the PageRank and CheiRank analysis recover about 75\% of
industrially developed countries of $G20$. However, now these countries
appear at the top ranking positions not due to their richness
but due to the efficiency of their trade network.
Another important aspect found in  \cite{wtngoogle}
is that both PageRank and CheiRank vectors appear
very naturally in the WTN corresponding to import and export flows.
 
In this work we extend the Google matrix analysis
for the multiproduct WTN obtained from COMTRADE \cite{comtrade}
with up to $N_p=61$ trade products for up to $N_c=227$ countries.
The global $G$ matrix of such trade flows has a size up to
$N=N_p N_c =13847$ nodes. The names and codes of products are given in 
Table~\ref{table1} and their trade volumes, expressed in percent of 
the whole world trade volume, are given in Table~\ref{table2}
for years 1998 and 2008.
The main problem of construction
of such a matrix is not its size, which is rather modest
compared to those studied in \cite{arxivrmp}, but the necessity
to treat all countries on democratic grounds
and at the same time to treat trade products on the basis of their
trade volume. Indeed, the products cannot be considered on democratic grounds
since their contributions to economy are linked with their trade
volume. Thus, according to Table~\ref{table2}, in year 2008 the trade volume of 
{\it Petroleum and petroleum products} (code 33 in Table~\ref{table1}) 
is by a factor 300 larger than
those of  {\it Hides, skins and fur skins (undress.)} (code 21 in Table~\ref{table1}).
To incorporate these features in our mathematical analysis of
multiproduct WTN we developed in this work
the Google Personalized Vector Method (GPVM) which
allows to keep a democratic treatment of countries
and at the same time to consider products proportionally to
their trade volume. As a result we are able to perform 
analysis of the global multiproduct WTN 
keeping all interactions between all countries and all products.
This is a new step in the WTN analysis 
since in our previous studies \cite{wtngoogle}
it was possible to consider a trade between countries only in one product
or only in all products summed together (all commodities).
The new finding of such global WTN analysis is an asymmetric
ranking of products: some of them are more oriented to import
and others are oriented to export while the ranking of products
by the trade volume is always symmetric after summation over all countries.
This result with asymmetric ranking of products confirms the indications
obtained on the basis ecological ranking \cite{wtnecology}
which also give an asymmetry of products in respect to
import and export. Our approach also allows to analyze 
the sensitivity of trade network to price variations
of a certain product.

We think that the GPVM approach allows  to perform a most 
advanced analysis of multiproduct world trade.
The previous studies  have been restricted to 
studies of statistical characteristics of WTN links, patterns and their topology
(see e.g. 
\cite{garlaschelli,vespignani,fagiolo1,hedeem,fagiolo2,garlaschelli2010,fagiolo3}).
The applications of PageRank algorithm to the WTN 
was discussed in \cite{benedictis}, the approach based on HITS algorithm
was used in \cite{plosjapan}. In comparison to the above studies, 
the approach developed here
for the multiproduct WTN has an advantage of analysis of ingoing and outgoing flows, 
related to PageRank and CheiRank,  and of taking into account of multiproduct aspects
of the WTN.  Even if the importance to multiproduct WTN analysis 
is clearly understood by
researchers (see e.g. \cite{hidalgo}) the Google matrix methods
have not been efficiently used up to now. 
We also note that the matrix methods are extensively used for analysis
of correlations of trade indexes (see e.g. \cite{bouchaud,guhr})
but these matrices are Hermitian being qualitatively different from 
those appearing in the frame of Markov chains.
Here we make the steps in multi-functional
or multiproduct Google matrix analysis of the WTN extending the approach 
used in \cite{wtngoogle}.

\section{Methods}

\subsection{Google matrix construction for the WTN}

For a given year, we build $N_p$ money matrices $M^p_{c,c^\prime}$ of 
the WTN from the COMTRADE database \cite{comtrade} (see \cite{wtngoogle}).
\begin{equation}
M^p_{c,c^\prime} = \text{product $p$ transfer (in USD)
from country $c^\prime$ to $c$}
\label{eq1}
\end{equation}
Here the country indexes are $c,c^\prime=1,\ldots,N_c$ and 
a product index is $p=1,\ldots,N_p$. 
According to the COMTRADE database  
the number of UN registered countries is $N_c=227$ (in recent years) 
and the number of products is 
$N_p=10$ and $N_p=61$ for 1 and 2 digits respectively from the
Standard International Trade Classification (SITC) Rev. 1.
For convenience of future notation we also define the volume of imports and exports 
for a given country and product respectively as
\begin{equation}
V^p_c=\sum_{c^\prime} M^p_{c,c^\prime} \, , \,\;
V^{*p}_c=\sum_{c^\prime} M^p_{c^\prime,c} .
\label{eq2}
\end{equation}
The import and export volumes $V_c=\sum_p V^p_c$ and
$V^*_c = \sum_p V^{*p}_c$ are shown for the world map of countries in Fig.~\ref{fig1}
for year 2008.

In order to compare later with PageRank and CheiRank probabilities 
we define volume trade ranks in
the whole trade space of dimension $N=N_p\times N_c$. Thus the
ImportRank ($\hat{P}$) and ExportRank ($\hat{P}^*$) probabilities
are given by the normalized import and export volumes
\begin{equation}
\hat{P}_{i} = {V^p_c}/{V} \, , \,\;
\hat{P}^*_{i} = {V^{*p}_c}/{V} \, ,
\label{eq3}
\end{equation}
where $i=p+(c-1)N_p$, $i=1,\ldots,N$ and the total trade volume is
$V=\sum_{p,c,c^\prime} M^p_{c,c^\prime}=\sum_{p,c}V^p_c=\sum_{p,c}V^{*p}_c$.

The Google matrices $G$ and $G^*$ are defined as $N\times N$ real
 matrices with non-negative elements:
\begin{equation}
G_{ij}= \alpha S_{ij}+(1-\alpha) v_i e_j \, ,\; 
{G^*}_{ij}=\alpha {S^*}_{ij}+(1-\alpha) v^*_i e_j \, ,
\label{eq4}
\end{equation}
where $N=N_p\times N_c$, $\alpha \in (0,1]$ is the damping factor ($0<\alpha<1$), 
$e_j$ is the row 
vector of unit elements ($e_j=1$), and $v_i$ is a 
positive column vector called a \emph{personalization vector} 
with $\sum_i v_i=1$ \cite{meyer}.
We note that the usual Google matrix is recovered 
for a personalization vector $v_i=e_i/N$ 
In this work, following \cite{wtngoogle}, we fix $\alpha=0.5$. As discussed in 
\cite{meyer,arxivrmp,wtngoogle} a variation of $\alpha$ in a range $(0.5,0.9)$
does not significantly affect the probability distributions of PageRank 
and CheiRank vectors. We specify the choice of the personalization vector
a bit below.

The matrices $S$ and $S^*$  are built from money matrices ${M^p}_{cc'}$ as
\begin{eqnarray}
 \nonumber
S_{i,i^\prime}&=&\left\{\begin{array}{cl}   
M^p_{c,c^\prime} \delta_{p,p^\prime}/V^{*p}_{c^\prime}& 
\text{    if } V^{*p}_{c^\prime}\ne0\\ 
1/N & \text{    if } V^{*p}_{c^\prime}=0\\ 
\end{array}\right.\\
S^*_{i,i^\prime}&=&\left\{\begin{array}{cl}   
M^p_{c^\prime,c} \delta_{p,p^\prime}/V^{p}_{c^\prime}& 
\text{    if } V^{p}_{c^\prime}\ne0\\ 
1/N & \text{    if } V^{p}_{c^\prime}=0\\ 
\end{array}\right.
\label{eq5}
\end{eqnarray}
where $c,c^\prime=1,\ldots,N_c$; $p,p^\prime=1,\ldots,N_p$; 
$i=p+(c-1)N_p$; $i^\prime=p^\prime+(c^\prime-1)N_p$; 
and therefore $i,i^\prime=1,\ldots,N$. Note that the sum of each column of 
$S$ and $S^*$ are normalized to unity and hence the matrices $G, G^*, S, S^*$
belong to the class of Google matrices and Markov chains. 
The eigenvalues and eigenstates of $G, G^*$ are obtained by a direct numerical
diagonalization using the standard numerical packages.

\subsection{PageRank and CheiRank vectors from GPVM}

PageRank and CheiRank ($P$ and $P^*$) are defined as the right eigenvectors of
$G$ and $G^*$ matrices respectively at eigenvalue $\lambda=1$:
\begin{eqnarray}
\sum_j G_{ij} \psi_j= \lambda \psi_i \, , \; 
\sum_j {G^*}_{ij} {\psi^*}_j = \lambda {\psi^*}_j \; .
\label{eq6}
\end{eqnarray}
For the eigenstate at $\lambda=1$ we use the notation
$P_i=\psi_i , P^*={\psi^*}_i$ with the normalization 
$\sum P_i = \sum_i {P^*}_i=1$. For other eigenstates we use
the normalization $\sum_i |\psi_i|^2=\sum_i |\psi^*_i|^2=1$.
According to the Perron-Frobenius theorem
the components of $P_i$, ${P^*}_i$ are positive and give 
the probabilities
to find a random surfer on a given node \cite{meyer}.
The PageRank $K$ and CheiRank $K^*$ indexes
are defined from the decreasing ordering of $P$ and $P^*$ as
$P(K)\ge P(K+1)$ and $P^*(K)\ge P^*(K^*+1)$ with $K,K^*=1,\ldots,N$. 

If we want to compute the reduced PageRank and CheiRank probabilities of countries 
for \emph{all commodities} (or equivalently all products) 
we trace over the product space getting 
$P_c=\sum_{p} P_{pc}=\sum_{p}P\left(p+(c-1)N_p\right)$ 
and $P^*_c= \sum P^*_{pc}=\sum_{p}P^*\left(p+(c-1)N_p\right)$ 
with their corresponding $K_c$ and $K^*_c$ indexes. 
In a similar way we obtain the reduced PageRank and CheiRank probabilities
for products tracing over all countries and getting\\
$P_p=\sum_{c}P\left(p+(c-1)N_p\right) \sum_{p} P_{pc}$ and \\
$P^*_p=\sum_{c}P^*\left(p+(c-1)N_p\right) \sum P^*_{pc} $ 
with their corresponding product indexes $K_p$ and $K^*_p$.

In summary we have $K_p,K^*_p=1,\ldots,N_p$ and 
$K_{c},K^*_{c}=1,\ldots,N_c$. A similar definition of ranks 
from import and export trade volume
can be done 
in a straightforward way via probabilities
$\hat{P}_p,\hat{P}^*_p,\hat{P}_c,\hat{P}^*_c,\hat{P}_{pc},\hat{P}^*_{pc}$ and 
corresponding indexes
$\hat{K}_p,\hat{K}^*_p,\hat{K}_c,\hat{K}^*_c,\hat{K},\hat{K}^*$.

\begin{figure}[!ht]
\begin{center}
\includegraphics[width=0.47\textwidth]{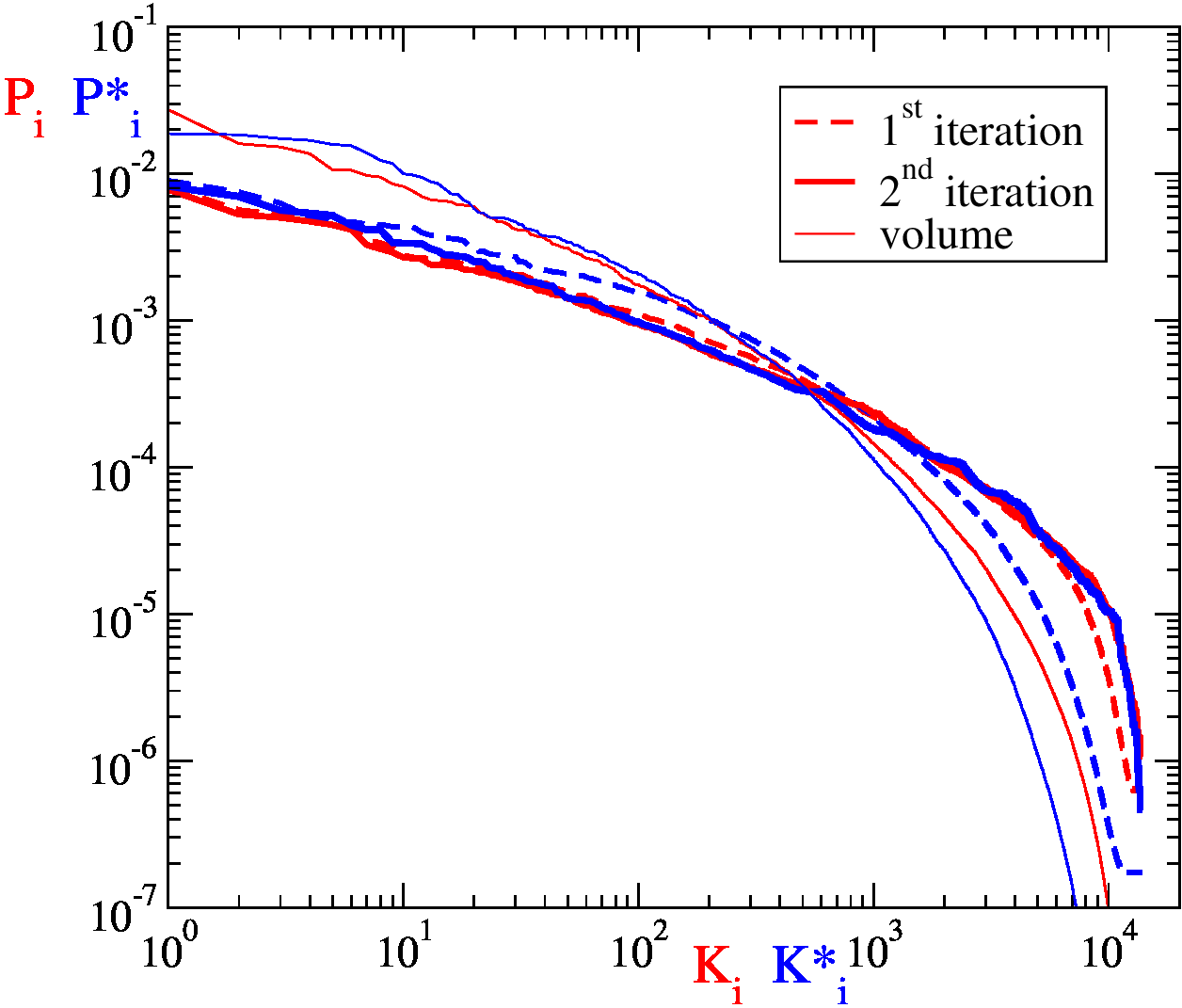} 
\end{center}
\vglue -0.1cm
\caption{
Dependence of probabilities of PageRank $P(K)$, CheiRank $P^*(K^*)$, 
ImportRank $\hat{P}(\hat{K})$ and ExportRank $\hat{P}^*(\hat{K}^*)$ 
as a function of their indexes in logarithmic scale 
for WTN in 2008 with $\alpha=0.5$
at $N_c=227$, $N_p=1$, $N=13847$. Here the results for GPVM after
$1^{\text{st}}$ and $2^{\text{nd}}$ iterations are shown 
for PageRank (CheiRank) in red (blue)
with dashed and solid curves respectively. 
ImportRank and ExportRank 
(trade volume) are shown by red and blue thin curves respectively. 
The fit exponents for PageRank and CheiRank are $\beta=0.61,0.7$
for the first iteration, $\beta=0.59,0.65$ for the second iteration, 
and $\beta=0.94,1.04$ for 
ImportRank and ExportRank (for the range $K\in[10,2000]$).
}
\label{fig2}
\end{figure}

\begin{figure}[!ht]
\begin{center}
\includegraphics[width=0.47\textwidth]{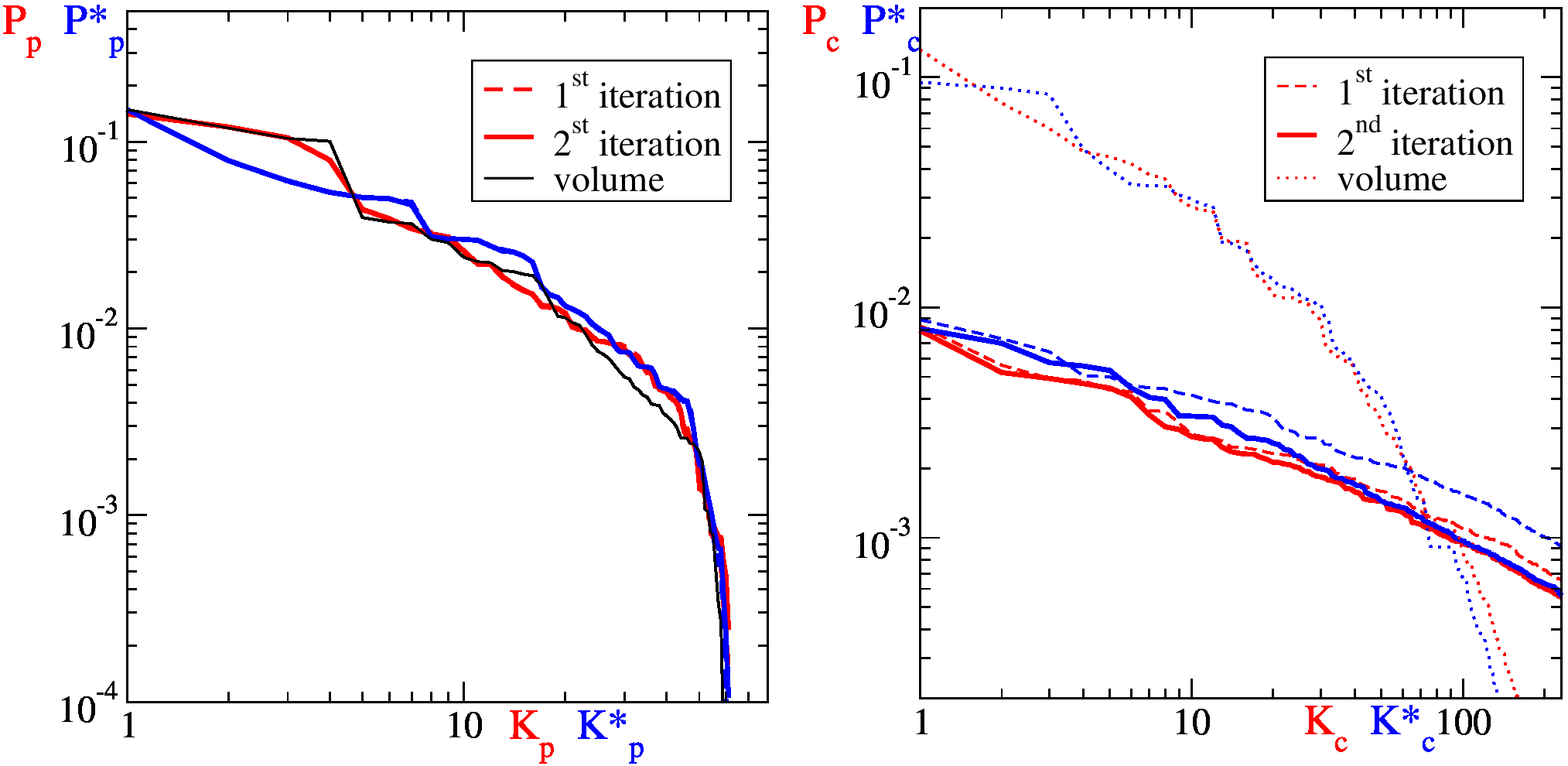} 
\end{center}
\vglue -0.1cm
\caption{
Probability distributions of PageRank and CheiRank for products $P_p(K_p)$,
$P^*_p(K^*_p)$ (left panel) 
and countries $P_c(K_c)$, $P^*_c(K^*_c)$ (right panel) 
in logarithmic scale for WTN from Fig.~\ref{fig2}. Here the results for the
$1^{\text{st}}$ and $2^{\text{nd}}$ GPVM iterations are shown 
by red (blue) curves for PageRank (CheiRank) 
with dashed and solid curves respectively. The probabilities from the 
trade volume ranking are shown by black curve (left) and 
dotted red and blue curves (right)
for ImportRank and ExportRank respectively.
}
\label{fig3}
\end{figure}

To compute the PageRank and CheiRank probabilities
from $G$ and $G^*$ keeping democracy in countries and 
proportionality of products to their trade volume we use the GPVM approach with
a personalized vector in (\ref{eq4}).
At the first iteration of Google matrix we  take into account 
the relative product volume per country using 
the following personalization vectors for $G$ and $G^*$:
\begin{equation}
v_i = \frac{V^p_c}{N_c \sum_{p^\prime} V^{p^\prime}_c} \, , \;
v^*_i = \frac{V^{*p}_c}{N_c \sum_{p^\prime} V^{*p^\prime}_c} \, ,
\label{eq7}
\end{equation}
using the definitions (\ref{eq2}) and the relation
$i=p+(c-1)N_p$.
This personalized vector depends both on product and country indexes.
In order to have the same value of personalization vector in countries 
we can define the second iteration vector proportional to the reduced 
PageRank and CheiRank vectors in products obtained from the 
GPVM Google matrix of the first iteration: 
\begin{equation}
v^\prime(i) = \frac{P_p}{N_c} \, , \;
v^{\prime *}(i) = \frac{P^*_p}{N_c} \, .
\label{eq8} 
\end{equation}
In this way we keep democracy in countries but weighted products.
This second iteration personalized vectors are used for the main part of
computations and operations with $G$ and $G^*$. This procedure with two iterations
forms  our GPVM approach.
The difference between results obtained from
the first and second iterations is not very large 
(see Figs.~\ref{fig2},~\ref{fig3})
but a detailed analysis of ranking of countries and products 
shows that the personalized vector for the second iteration
improves the results making them more stable and less fluctuating.
In all Figures below (except Figs.~\ref{fig2},~\ref{fig3})
we show the results after the second iteration.

The obtained results show the distribution of nodes on the 
PageRank-CheiRank plane $(K,K^*)$. 
In addition to two ranking indexes $K,K^*$
we use also 2DRank index $K_2$ which combines
the contribution of these indexes 
as described in \cite{wikizzs}.
The ranking list   $K_2(i)$ is constructed by  
increasing $K \rightarrow K+1$ 
and increasing 2DRank index 
$K_2(i)$ by one if a new entry is present in the list of
first $K^*<K$ entries of CheiRank, then the one unit step is done in
$K^*$ and $K_2$ is increased by one if the new entry is
present in the list of first $K<K^*$ entries of CheiRank.
More formally, 2DRank $K_2(i)$ gives the ordering 
of the sequence of sites, that $\;$ appear
inside  $\;$ the squares  $\;$
$\left[ 1, 1; \;K = k, K^{\ast} = k; \; \-... \right]$ when one runs
progressively from $k = 1$ to $N$.
Additionally, we analyze the distribution of nodes
for reduced indexes $(K_p,K^*_p)$, $(K_c,K^*_c)$.

We also characterize the localization properties of eigenstates of $G, G^*$
by the inverse participation ration (IPR)
defined as $\xi = (\sum_i |\psi_i|^2)^2/\sum_i |\psi_i|^4$. This
characteristic determines an effective 
number of nodes which contribute to a formation
of a given eigenstate (see details in \cite{arxivrmp}). 

\subsection{Correlators of PageRank and CheiRank vectors}

Following previous works \cite{linux,wikizzs,wtngoogle}
the correlator of PageRank and CheiRank vectors is defined as:
\begin{equation}
\kappa=N \sum_{i=1}^{N} P(i) P^*(i) - 1  \; .
\label{eq9} 
\end{equation}
The typical values of $\kappa$ are given in \cite{arxivrmp}
for various networks.

For global PageRank and CheiRank the product-product 
correlator matrix is defined as: 
\begin{footnotesize}
\begin{equation}
\kappa_{p p^\prime}=N_c\sum_{c=1}^{N_c}\left[\frac{P(p+(c-1)N_p)P^*(p^\prime+(c-1)N_p)}
{\sum_{c^\prime} P(p+(c^\prime-1)N_p) \sum_{c^{\prime\prime}} 
P^*(p^{\prime}+(c^{\prime\prime}-1)N_p)}\right] -1
\label{eq10} 
\end{equation}
\end{footnotesize}

Then the correlator for a given product is obtained from (\ref{eq10}) as:
\begin{equation}
\kappa_{p}=\kappa_{p p^\prime} \delta_{p,p^{\prime}} \, , 
  \label{eq11} 
\end{equation}
where $\delta_{p,p^{\prime}}$ is the Kronecker delta.

We also use the correlators obtained from the 
probabilities traced over products 
($P_c=\sum_p P_{pc}$) and over countries
($P_p=\sum_c P_{pc}$)  which are defined as
\begin{equation}
\kappa(c)=N_c \sum_{c=1}^{N_c} P_{c} P^*_{c} - 1  \, , \; 
\kappa(p)=N_p \sum_{p=1}^{N_p} P_{p} P^*_{p} - 1  \, .
\label{eq12} 
\end{equation}

In the above equations (\ref{eq9})-(\ref{eq12})
the correlators are computed for PageRank and CheiRank probabilities.
We can also compute the same correlators using probabilities from the trade
volume in ImportRank $\hat{P}$ and ExportRank $\hat{P}^*$ defined 
by (\ref{eq3}).

We discuss the values of these correlators in Section 4.

\section{Data description}
All data are obtained from the COMTRADE database \cite{comtrade}.
We used products from COMTRADE SITC Rev. 1 classification 
with number of products $N_p=10$ and $61$. We choose SITC Rev.1
since it covers the longest time interval. The main results are
presented for $N_p=61$ with up to $N_c=227$ countries.
The names of products are given in Table~\ref{table1},
their ImportRank index $K$ and their fraction (in percent) of global trade
volume in years 1998 and 2008 are given in Table~\ref{table2}.
The data are collected and presented for the years 1962 - 2010.
Our data and results are available at 
\cite{ourwebpage}, the data for the matrices
$M^p_{c,c^{\prime}}$ are available at  COMTRADE \cite{comtrade}
with the rules of their distribution policy.
Following \cite{wtngoogle} we use for countries ISO 3166-1 alpha-3 code
available at Wikipedia.

\section{Results}

We apply the above methods to the described data sets of COMTRADE
and present the obtained results below.

\subsection{PageRank and CheiRank probabilities}


The dependence of probabilities of PageRank $P(K)$ and CheiRank $P^*(K^*)$
vectors on their indexes $K, K^*$ are shown in Fig.~\ref{fig2}
for a selected year 2008. The results can be approximately described by an
algebraic dependence $P \propto 1/K^\beta$ with the exponent values
given in the caption. It is interesting to note that 
we find approximately the same $\beta \approx 0.6$ both for
PageRank and CheiRank in contrast to the WWW, universities and Wikipedia
networks where usually one finds $\beta \approx 1$ for PageRank and 
$\beta \approx 0.6$ for CheiRank \cite{meyer,arxivrmp}.
We attribute this to an intrinsic property of WTN
where the countries try to keep economy balance of their trade.
The data show that the range of
probability variation is reduced for the Google ranking compared to
the volume ranking. This results from a democratic ranking 
of countries used in the Google matrix analysis that 
gives a reduction of richness dispersion between countries.
The results also show that the variation of probabilities 
for 1st and 2nd GPVM results are not very large 
that demonstrates the convergence of this approach.

After tracing probabilities over countries we obtain 
probability distributions $P_p(K_p)$, $P^*_p(K^*_p)$ 
over products shown in Fig.~\ref{fig3}. The variation range
of probabilities is the same as for the case of volume ranking.
This shows that the GPVM approach correctly treats products
keeping their contributions proportional to their volume.
The difference between 1st and 2nd iterations is rather small
and is practically not visible on this plot.
The important result well visible here is
a visible difference between PageRank and CheiRank
probabilities while there is no difference
between ImportRank and ExportRank probabilities
since they are equal after tracing over countries.

After tracing over products we obtain probability
distributions $P_c(K_c)$, $P^*_c(K^*_c)$ over countries shown in Fig.~\ref{fig3}. 
We see that the probability of volume ranking
varies approximately by a factor $1000$ while for
PageRank and CheiRank such a factor is only approximate $10$.
Thus the democracy in countries induced by the Google matrix construction
reduces significantly the variations of probabilities among countries
and inequality between countries.

Both panels of Fig.~\ref{fig3} show relatively small variations
between 1st and 2nd GPVM iterations confirming the stability
of this approach. In next sections we present the results
only for 2nd GPVM iteration. This choice is confirmed by consideration
of ranking positions of various nodes of global matrices $G, G^*$
which show less fluctuations compared to the results of the 1st GPVM iteration.

From the global ranking of countries and products we can select
a given product and then determine local
ranking of countries in a given product to see how strong 
is their trade for this product. The results for 
three selected products are discussed below for year
2008. For comparison we also present comparison with the export-import ranking 
from the trade volume.

\begin{figure}[!ht]
\begin{center}
\includegraphics[width=0.47\textwidth]{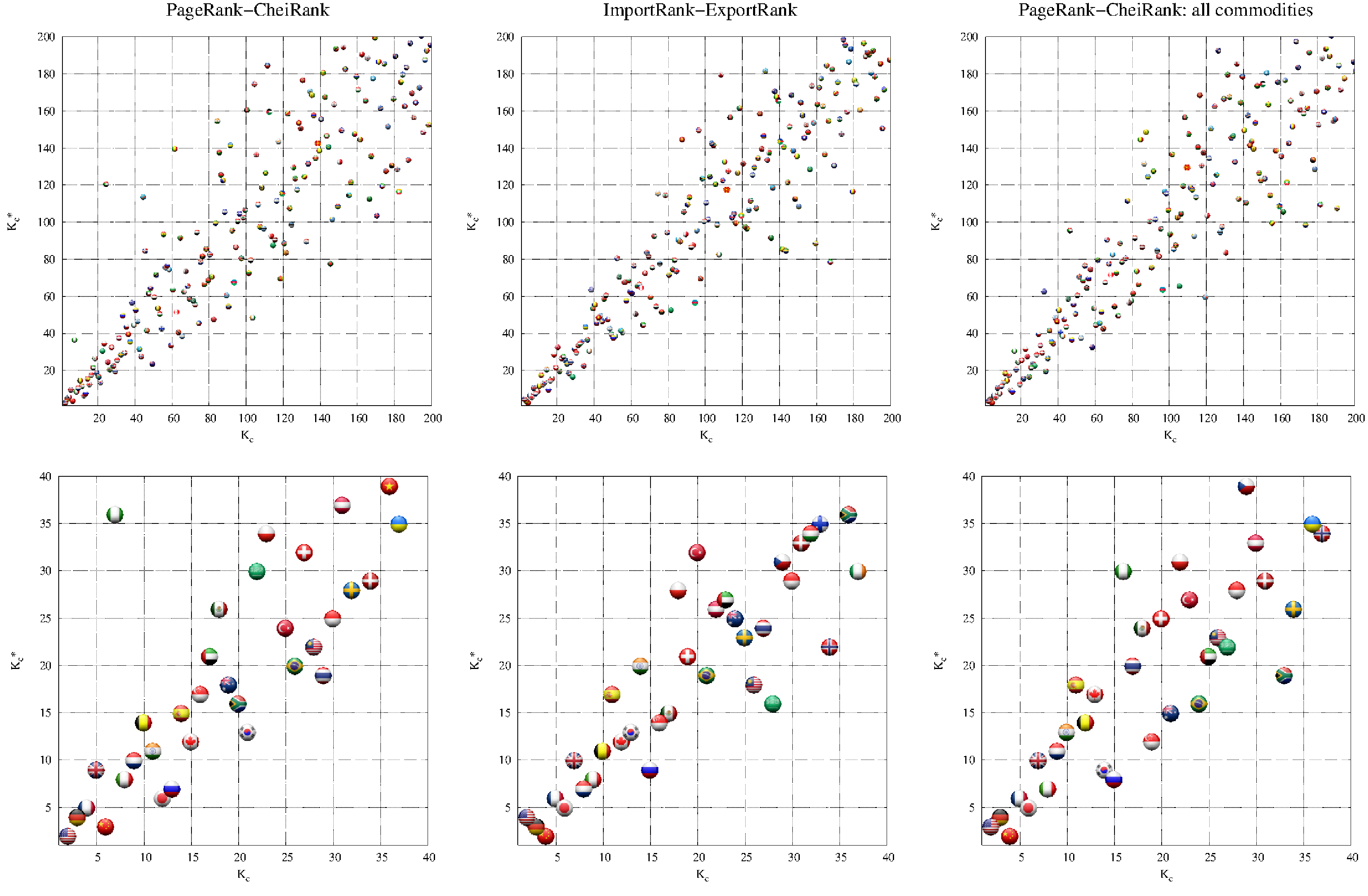} 
\end{center}
\vglue -0.1cm
\caption{
Country positions on PageRank-CheiRank plane $(K_c,K^*_c)$ 
obtained by the GPVM analysis (left panels),
ImportRank-ExportRank of trade volume (center panels), and 
for PageRank-CheiRank of \emph{all commodities} 
(right panels, data from \cite{wtngoogle}).
Top panels show global scale ($K_c,K^*_c\in[1,200]$) 
and bottom panels show zoom on top ranks ($K_c,K^*_c\in[1,40]$).
Each country is shown by circle with its own flag 
(for a better visibility the circle center 
is slightly displaced from its integer position $(K_c,K^*_c)$ 
along direction angle $\pi/4$). Data are shown for year 2008.}
\label{fig4}
\end{figure}

\subsection{Ranking of countries and products}

After tracing the probabilities $P(K), P^*(K^*)$ 
over products we obtain the distribution of world countries on the
PageRank-CheiRank plane $(K_c,K^*_c)$ presented in Fig.~\ref{fig4} 
for a test year 2008. In the same figure we present the rank distributions 
obtained from ImportRank-ExportRank probabilities of trade volume 
and the results obtained in \cite{wtngoogle} for trade in {\it all commodities}.
For the GPVM data we see the global features already discussed in 
\cite{wtngoogle}: the countries are distributed in a vicinity of diagonal
$K_c=K^*_c$ since each country aims to keep its trade balanced.
The top $20$ list of top $K_2$ countries recover $15$ of 19 countries of
$G20$  major world economies (EU is the number 20) 
thus obtaining 79\% of the whole list.
This is close to the percent obtained in \cite{wtngoogle}
for trade in {\it all commodities}. 

The global distributions 
of top countries with $K_c \leq 40$, $K^*_c \leq 40$
for the three ranking methods, shown in Fig.~\ref{fig4},
are similar on average. But some modifications introduced
by the GPVM analysis are visible. Thus China (CHN) moves on 2nd position 
of CheiRank while it is in the 1st position for trade volume ranking and
CheiRank  of {\it all commodities}. Also e.g. Saudi Arabia (SAU) and Russia (RUS)
move from the CheiRank positions ${K^*}_c=21$ and 
${K^*}_c=7$ in  {\it all commodities}  \cite{wtngoogle}
to ${K^*}_c=29$ and ${K^*}_c=6$ in the GPVM ranking, respectively. 
Other example is a significant displacement of Nigeria (NGA).
We explain such differences as the result of 
larger connectivity required for 
getting high ranking in the multiproduct WTN.
Indeed, China is more specialized in specific products
compared to USA
(e.g. no petroleum production and export)
that leads to its displacement in ${K^*}_c$.
We note that the ecological ranking 
gives also worse ranking positions for China
comparing to the trade volume ranking \cite{wtnecology}.
In a similar way the trade of Saudi Arabia 
is strongly dominated by petroleum  and moreover
its petroleum trade is strongly oriented on USA
that makes its trade network concentrated on a few links while Russia
is improving its position in ${K^*}_c$ due to significant 
trade links with EU and Asia.

In global, the comparison of three ranks of countries shown in Fig.~\ref{fig4}
confirms that the GPVM analysis gives a reliable ranking of multiproduct WTN.
Thus we now try to obtain  new features of multiproduct WTN
using the GPVM approach.

\begin{figure}[!ht]
\begin{center}
\includegraphics[width=0.47\textwidth]{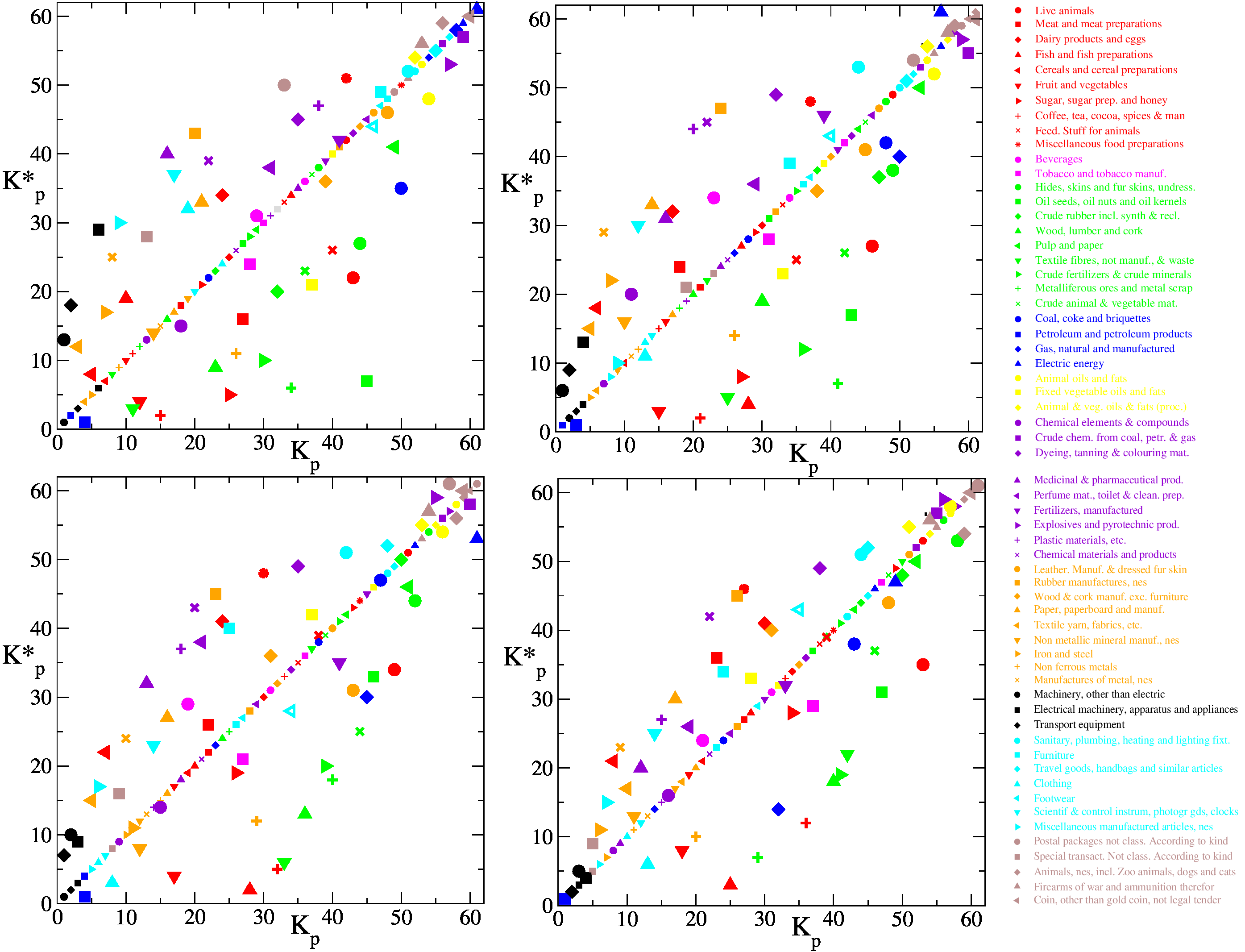} 
\end{center}
\vglue -0.1cm
\caption{
Two dimensional ranking of products on the 
PageRank-CheiRank plane $(K_p,K^*_p)$. 
Each product is represented by its specific  combination of color and symbol: 
color illustrates the first digit of COMTRADE SITC Rev. 1 code with 
the corresponding name shown in the legend on the right; symbols correspond to product names 
listed in Table~\ref{table1} with their code numbers; 
the order of names on the right panel of this Fig. is the same as in Table~\ref{table1}.
The trade volume ranking via ImportRank-ExportRank  
is shown by small symbols at the diagonal $\hat{K}_p=\hat{K}^*_p$,
after tracing over countries this ranking is symmetric in products.
Top left and right panels show  years 1963 and 1978, while bottom left and right 
panels show years 1993 and 2008 respectively.
}
\label{fig5}
\end{figure}

The main new feature obtained within the GPVM approach is shown in Fig.~\ref{fig5}
which gives the distribution of products on the PageRank-CheiRank plane
$(K_p, K^*_p)$ after tracing of global probabilities $P(K), P^*(K^*)$ over
all world countries. The data clearly show that the distribution of products
over this plane is asymmetric while the ranking of products from 
the trade volume produces the symmetric ranking of products
located directly on diagonal $K_p=K^*_p$. 
Thus the functions of products are asymmetric:
some of them are more oriented to export (e.g. {\it 03 Fish and fish 
preparations, 05 Fruit and vegetables,
26 Textile fibers, not manuf. etc., 28 Metalliferous ores and metal scrap,
84 Clothing});
in last years (e.g. 2008) {\it 34 Gas, natural and manufactured} also takes
well pronounced export oriented feature 
characterized by location in the lower right triangle $(K^*_p<K_p)$
of the square plane $(K_p,K^*_p)$. In contrast to that
the products located in the upper left triangle $(K^*_p>K_p)$
represent import oriented products
(e.g. 
{\it 02 Dairy products and eggs,  04 Cereals and cereal preparations,
64 Paper, paperboard and manuf., 65 Textile yarn, fabrics, etc.,
86 Scientific \& control instrum, photogr gds, clocks}).

It is interesting to note that the machinery products 71, 72. 73 are 
located on leading import oriented positions in 1963,  1978, 1993
but they become more close to symmetric positions
in 2008. We attribute this to development of China
that makes the trade in these products more symmetric
in import-export. It is interesting to note that in 1993 the
product {\it 33 Petroleum and petroleum products} 
loses its first trade volume position
due to low petroleum prices but still
it keeps the first CheiRank position 
showing its trade network importance for export.
Each product moves on $(K_p,K^*_p)$ with time.
However, a part of the above points, we can say that the global 
distribution does not manifest drastic changes. Indeed, e.g.
the green symbols of first digit 2 remain
export oriented 
for the whole period 1963  - 2008.
We note that the established asymmetry of products orientation
for the world trade is in agreement with the similar
indications obtained on the basis of ecological ranking in \cite{wtnecology}.
However, the GPVM approach used here have more solid mathematical 
and statistical foundations with a reduced significance of  fluctuations
comparing to the ecological ranking. 

\begin{figure}[!ht]
\begin{center}
\includegraphics[width=0.47\textwidth]{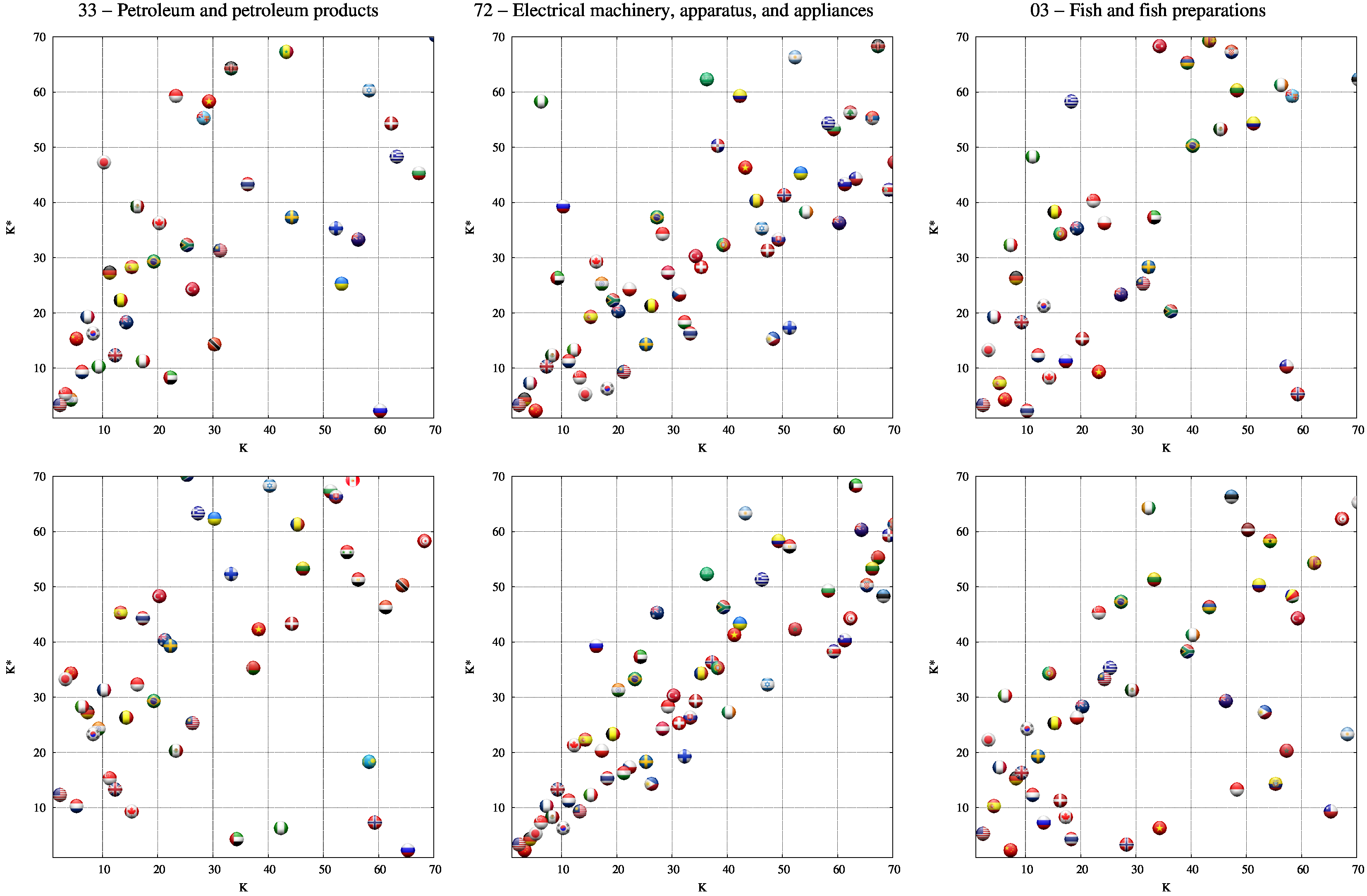} 
\end{center}
\vglue -0.1cm
\caption{
Top panels show results of the GPVM data for
country positions on PageRank-CheiRank plane 
of local rank values $K,K^*$ 
ordered by  $(K_{cp},K^*_{cp})$ for 
specific products with $p=33$ (left panel), $p=72$ (center panel)
and $p=03$ (right panel).
Bottom  panels show the   
ImportRank-ExportRank planes 
respectively for comparison.
Data are given for year 2008.
Each country is shown by circle with its own flag as in Fig.~\ref{fig4}.}
\label{fig6}
\end{figure}

The comparison between the GPVM and trade volume ranking methods 
provides interesting information. Thus in petroleum code 33
we have on top positions Russia, Saudi Arabia, United Arab Emirates
while from the  CheiRank order of this product we 
find  Russia, USA, India (see  Fig.~\ref{fig6} 
and Table~\ref{table3}). This marks the importance of the role
of USA and India played in the WTN 
and in the redistribution of petroleum over nearby 
region countries, e.g. around India. Also Singapore
is on a local petroleum position just behind India
and just before Saudi Arabia, see Table~\ref{table3}.
This happens due to strong involvement of India and Singapore
in the trade redistribution flows of petroleum
while Saudi Arabia has rather restricted trade connections
strongly oriented on USA and nearby countries.

For electrical machinery 72 there are less modifications
in the top export or CheiRank
positions (see Fig.~\ref{fig6})
but we observe significant broadening of positions on 
PageRank-CheiRank
plane comparing to ImportRank-ExportRank. Thus, Asian countries 
(China, Japan, S. Korea, Singapore)
are located on the PageRank-CheiRank plane
well below the 
diagonal $K=K^*$ showing a significant 
trade advantages of these countries 
in product 72 comparing to Western countries
(USA, Germany, France, UK).

Another product, shown in  Fig.~\ref{fig6},
is {\it 03 Fish and fish preparations}.
According to the trade volume export ranking the top three positions are
attributed to
China, Norway, Thailand. However, from CheiRank of product 03
we find another order with Thailand, USA, China.
This result stresses again the broadness and robustness of the
trade connections of Thailand and USA. As another example we note
a significant improvement of Spain CheiRank position 
showing its strong commercial relations for product 03.
On the other side Russia has relatively good position in 
the trade volume export of 03 product
but its CheiRank index becomes worse due to 
absence of broad commercial links for this product.

The global top 20 positions of indexes $K, K^*, K_2, {\hat K}, {\hat K^*}$
are given in Table~\ref{table3} for year 2008.
We note a significant improvement of positions
of Singapore and India in PageRank-CheiRank
positions comparing to their positions in the trade volume ranking.
This reflects their strong commercial relations in the world trade.
In the trade volume ranking the top positions are taken
by 33 petroleum and digit 7 of machinery products.
This remains mainly true for PageRank-CheiRank positions
but we see the spectacular improvement of 
positions of 84 Clothing for China ($K^*=2$)
and 93 Special transact. for USA ($K=4$)
showing thus these two products have 
strong commercial exchange all over the world
even if their trade volume is not so dominant.

\begin{figure}[!ht]
\begin{center}
\includegraphics[clip=true,width=0.47\textwidth]{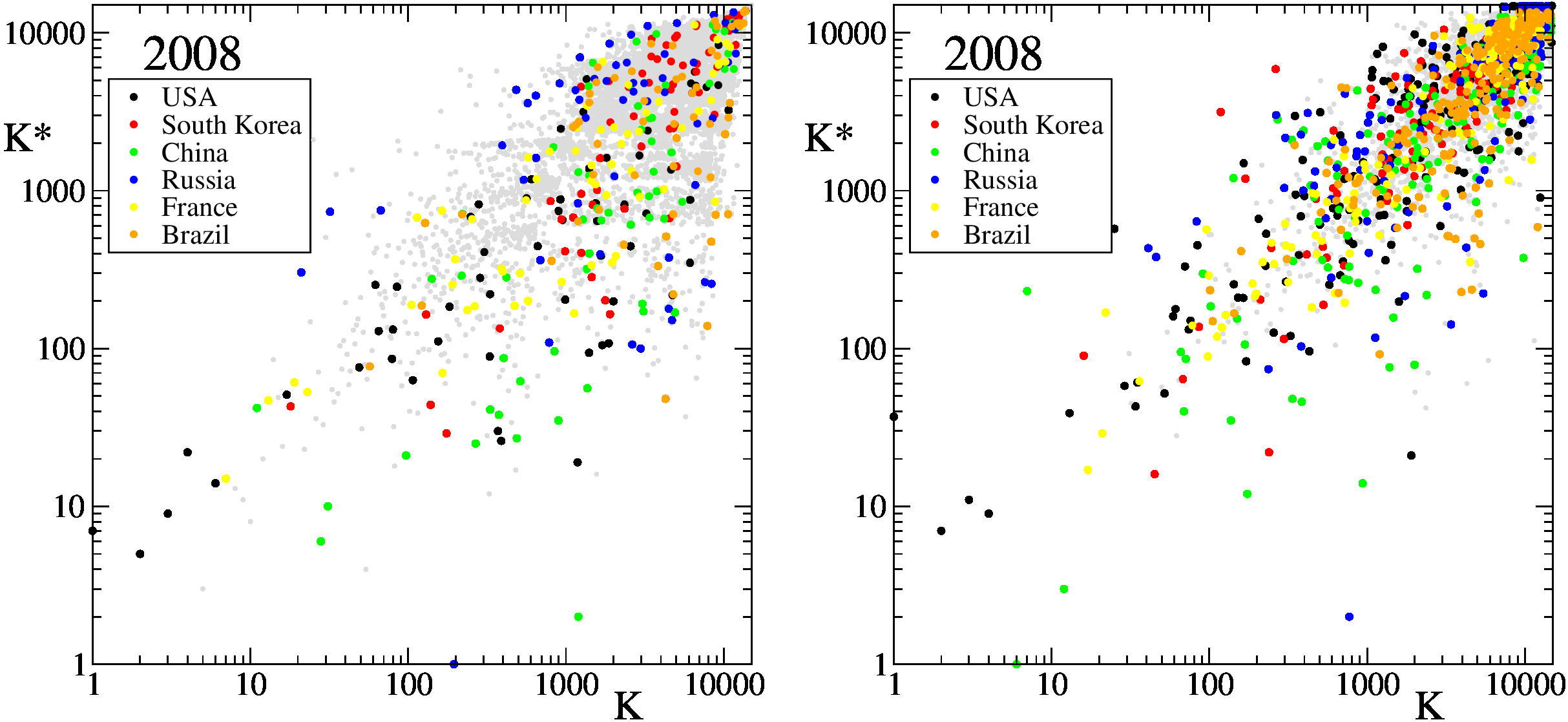} 
\end{center}
\vglue -0.1cm
\caption{Global plane of rank indexes $(K,K^*)$ for PageRank-CheiRank (left panel) 
and ImportRank-ExportRank (right panel) for $N=13847$ nodes in year 2008.
Each country and product pair is represented by a gray circle.
Some countries are highlighted in colors: USA with black, 
South Korea with red, China with green, Russia with red, 
France with yellow and Brazil with orange.}
\label{fig7}
\end{figure}

We show the plane $(K,K^*)$ for the global world ranking in 
logarithmic scale in 2008 in Fig.~\ref{fig7}. 
The positions of trade nodes of certain selected
countries are shown by color. We observe that the trade volume
gives a higher concentration of nodes around diagonal
comparing to the GPVM ranking. We attribute this 
to the symmetry of trade volume in products.

\begin{figure}[!ht]
\begin{center}
\includegraphics[width=0.47\textwidth]{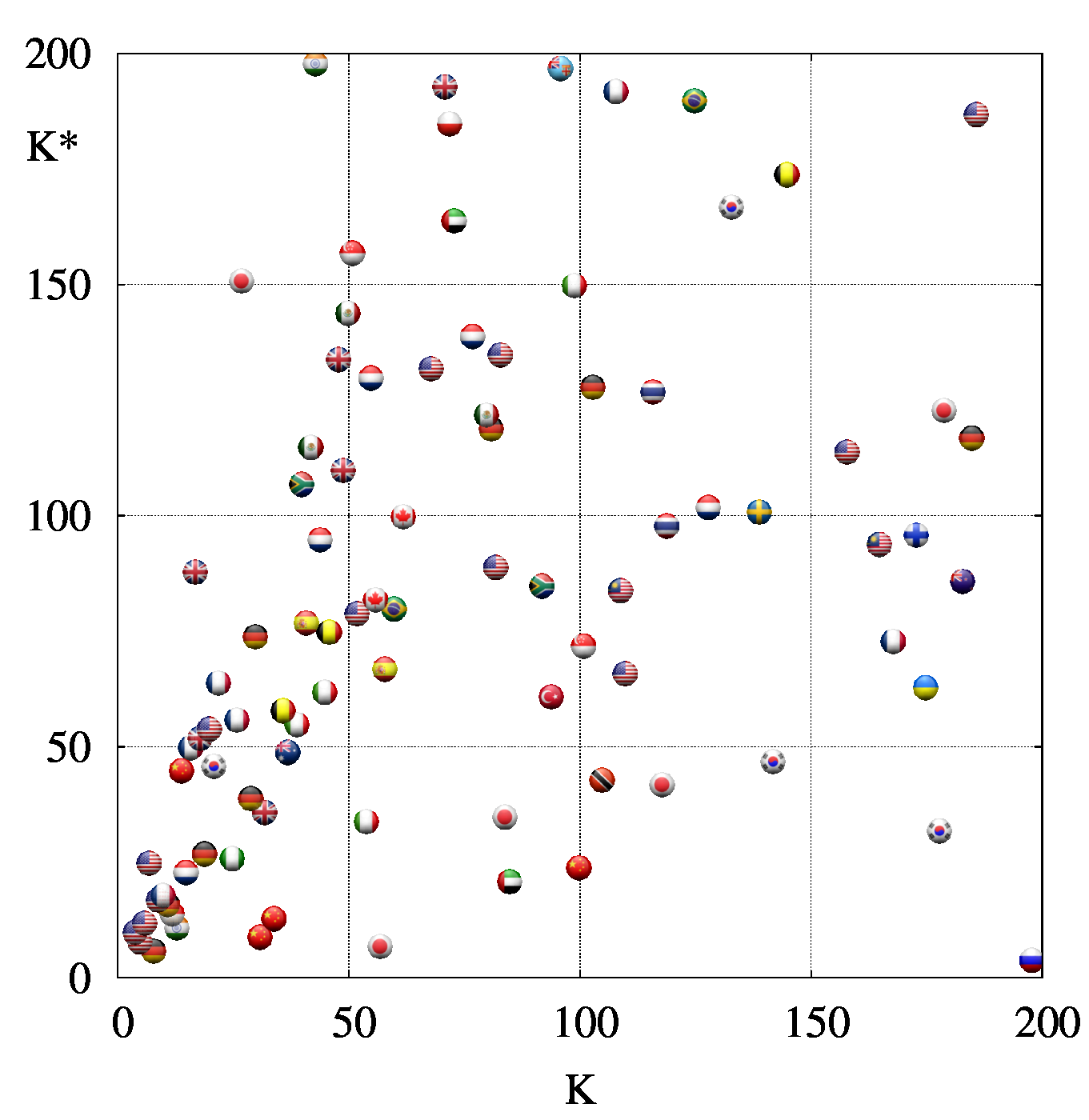}
 \end{center}
\vglue -0.1cm
\caption{Top 200 global PageRank-CheiRank indexes $(K,K^*)$ distributions for year 2008.
Each country (for different products) is represented by its flag.}
\label{fig8}
\end{figure}

In  Fig.~\ref{fig8} we show the distributions of top $200$ ranks of 
the PageRank-CheiRank plane (zoom of left panel of Fig.~\ref{fig7}).
Among the top 30 positions of $K^*$ there are
8 products of USA,  6  of China, 3 of Germany and other countries with less 
number of products. The top position at $K^*=1$ corresponds to product 33 of Russia
while Saudi Arabia is only at $K^*=12$ for this product. 
The lists of all $N=13847$ network nodes with their
$K,K_2,K^*$ values are available at \cite{ourwebpage}.

\subsection{Time evolution of ranking}

\begin{figure}[!ht]
\begin{center}
\includegraphics[width=0.47\textwidth]{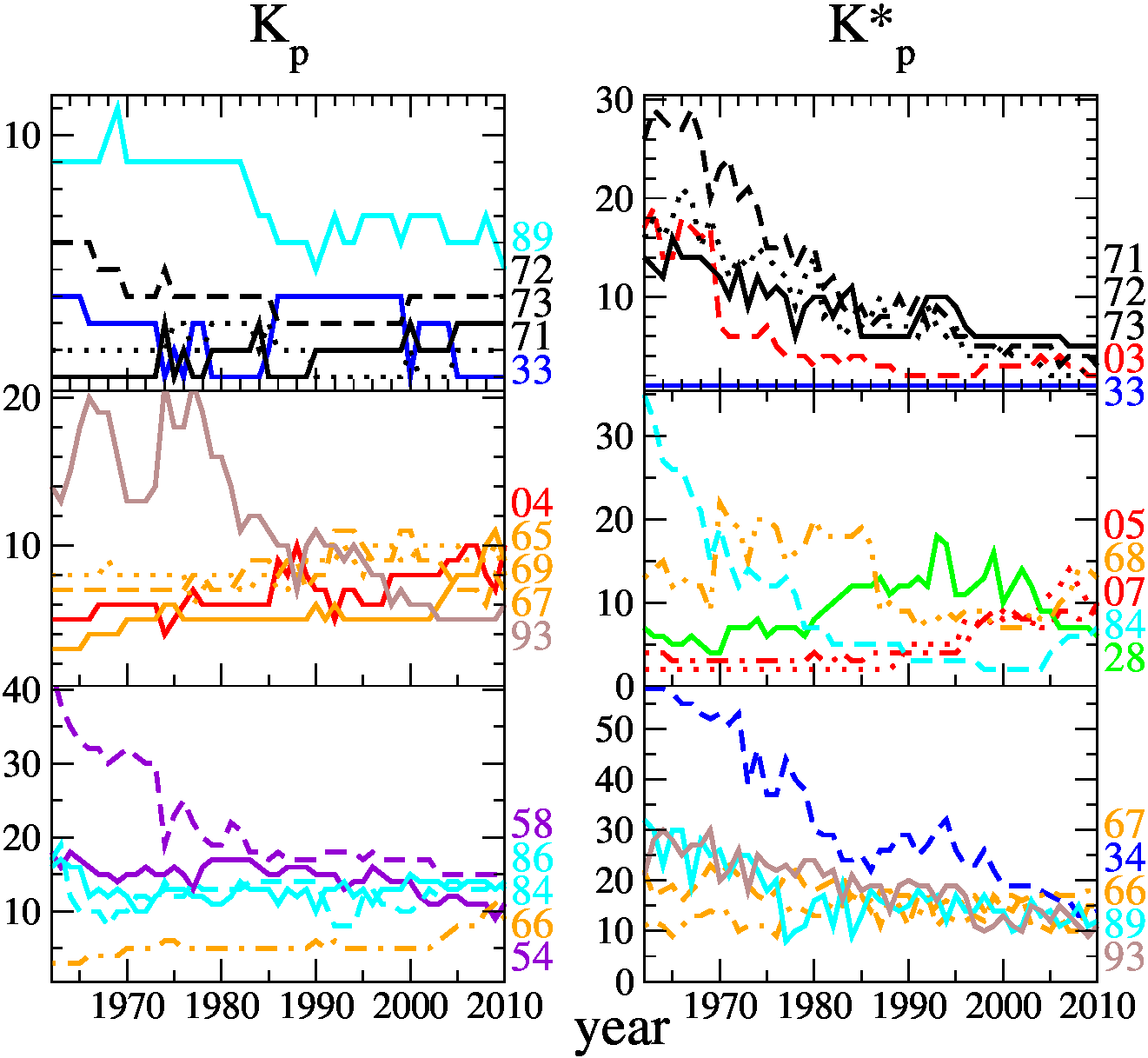} 
\end{center}
\vglue -0.1cm
\caption{
Time evolution of PageRank $K_p$ and CheiRank $K^*_p$
indexes for years 1962 to 2010 for certain products
marked on the right panel side by their codes from Table~\ref{table1}.
Top panels show top $5$ ranks of 2010,
middle and bottom panels show ranks 
$6$ to $10$ and $11$ to $15$ for 2010 respectively. Colors of curves 
correspond to the colors
of Fig.~\ref{fig5} marking the first code digit.}
\label{fig9}
\end{figure}

The time evolution of indexes of products $K_p, K^*_p$ is shown in Fig.~\ref{fig9}.
To obtain these data we trace PageRank and CheiRank probabilities over countries
and show the time evolution of rank indexes of products $K_p, K^*_p$ for
top 15 rank products of year 2010. The product {\it 33 Petroleum and petroleum products}
remains at the top CheiRank position $K^*_p=1$ for the whole period
while in PageRank it shows significant variations from $K_p=1$ to $4$
being at $K_p=4$ at 1986 - 1999 when the petroleum had a low price.
Products with first digit $7$ have high ranks of $K_p$ but especially strong variation
is observed for $K^*_p$ of {\it 72 Electrical machinery} moving from
position 26 in 1962 to 4 in 2010.
Among other indexes with strong variations we note {\it 58 Plastic materials},
{\it 84 Clothing}, {\it 93 Special transact.}, {\it 34 Gas, natural and manufactured}.

\begin{figure}[!ht]
\begin{center}
\includegraphics[width=0.47\textwidth]{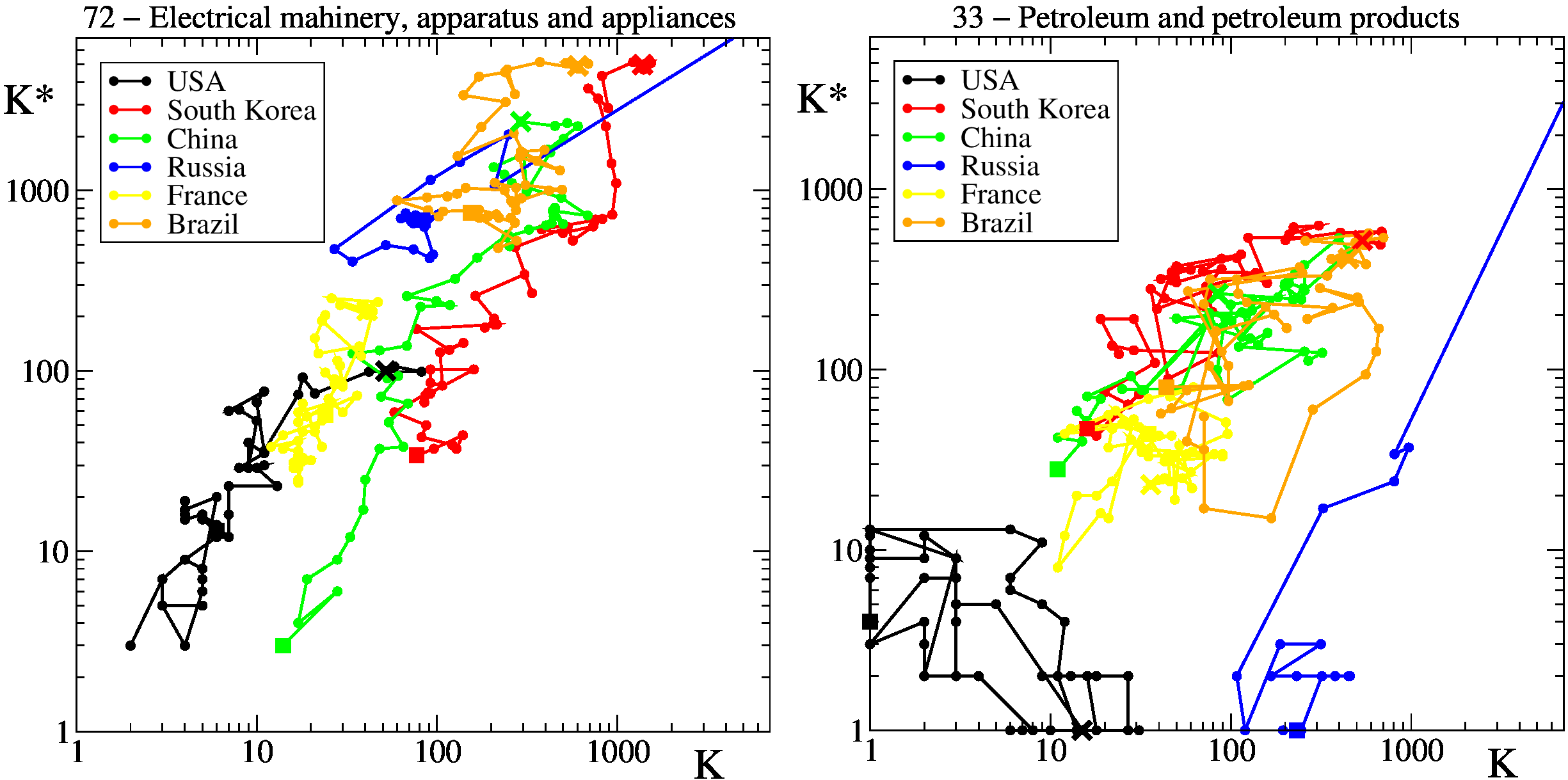} 
\end{center}
\vglue -0.1cm
\caption{
Time evolution of ranking of two products $72$ and $33$ 
for $6$ countries of Fig.~\ref{fig7} shown on the global
PageRank-CheiRank plane $(K,K^*)$.
Left and right panels show the cases of \emph{72 Electrical machinery, 
apparatus and appliances} and \emph{33 petroleum and petroleum products} respectively.
The evolution in time starts in 1962 (marked by cross) 
and ends in 2010 (marked by square).
}
\label{fig10}
\end{figure}

The time evolution of products $33$ and $72$ 
on the global index plane $(K,K^*)$
is shown in Fig.~\ref{fig10}
for 6 countries from Fig.~\ref{fig7}. Thus for product $72$
we see a striking improvement of $K^*$ for China and S.Korea
that is at the origin of the global importance improvement of 
$K^*_p$ in Fig.~\ref{fig9}. For the product $33$ in Fig.~\ref{fig10}
Russia improves significantly its rank positions
taking the top rank $K^*=1$ (see also Table~\ref{table3}).

\begin{figure}[!ht]
\begin{center}
\includegraphics[width=0.47\textwidth]{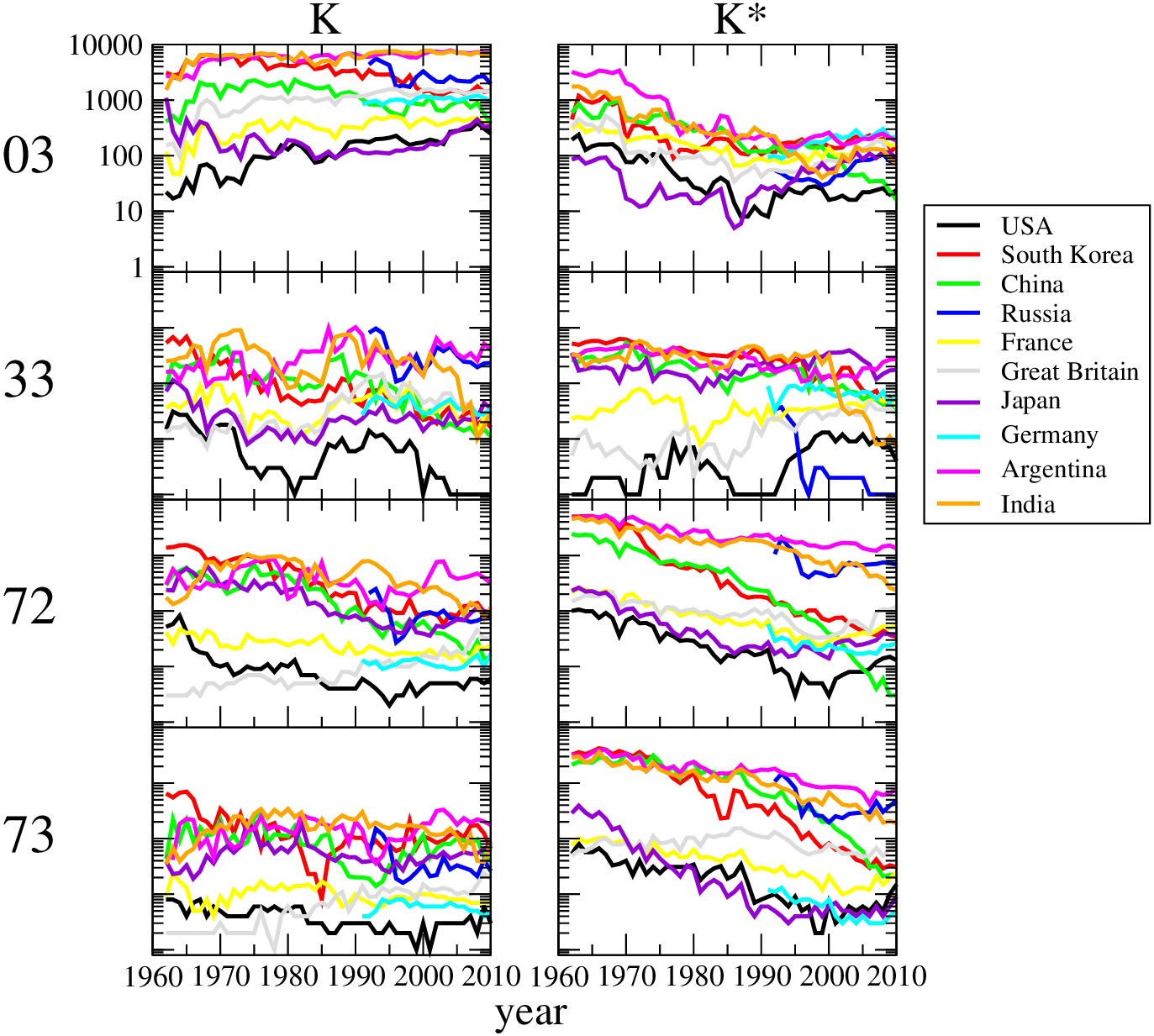} 
\end{center}
\vglue -0.1cm
\caption{
Time evolution of
global ranking of PageRank and CheiRank indexes $K, K^*$ 
for selected 10 countries and 4 products.
Left and right panels show $K$ and $K^*$ 
as a function of years for products: 
$03$ \emph{Fish and fish preparations}; 
$33$ \emph{Petroleum and petroleum products}; 
$72$, \emph{Electrical machinery, apparatus and appliances}; 
and $73$ \emph{Transport equipment} (from top to bottom).
In all panels the ranks are shown in logarithmic scale 
for 10 given countries: USA, South Korea, 
China, Russia, France, Brazil, Great Britain, Japan, Germany and Argentina
marked by curve colors.}
\label{fig11}
\end{figure}

The variation of global ranks $K$, $K^*$ with time is shown for 4 products
and 10 countries in Fig.~\ref{fig11}. For products $72, 73$ on a scale of 50 years
we see a spectacular improvement of $K^*$ for China, Japan, S.Korea.
For the product $33$ we see strong improvement of $K^*$ for Russia
in last 15 years. It is interesting to note
that at the period 1986-1992 of cheap petroleum $33$ USA
takes the top position $K^*=1$ with a significant increase of its corresponding $K$
value.
We think that this is a result of political decision to make an
economical pressure on USSR since  such an increase of export of 
cheap price petroleum is not justified from the economical view point.
For the product $33$ we also note a notable improvement of $K^*$ of India
which is visible in CheiRank but not in ExportRank (see Table~\ref{table3}).
We attribute this not to a large amount of trade volume but to a significant
structural improvements of trade network of India in this product.
We note that the strength and efficiency of trade network is 
also at the origin of significant
improvement of PageRank and CheiRank positions of Singapore
comparing to the trade volume ranking.
Thus the development of trade connections of certain countries
significantly improves their Google rank positions. 
For the product $03$ we note the improvement of $K^*$
positions of China and Argentina while Russia shows no improvements in 
this product trade for this time period.

\subsection{Correlation properties of PageRank and CheiRank}

The properties of $\kappa$ correlator of PageRank and CheiRank 
vectors for various networks are reported in \cite{linux,arxivrmp}. 
There are directed networks with small or even slightly negative
values of $\kappa$, e.g. Linux Kernel or Physical Review citation networks,
or with $\kappa \sim 4$ for Wikipedia networks
and even larger values $\kappa \approx 116$ for the Twitter network.

\begin{figure}[!ht]
\begin{center}
\includegraphics[width=0.47\textwidth]{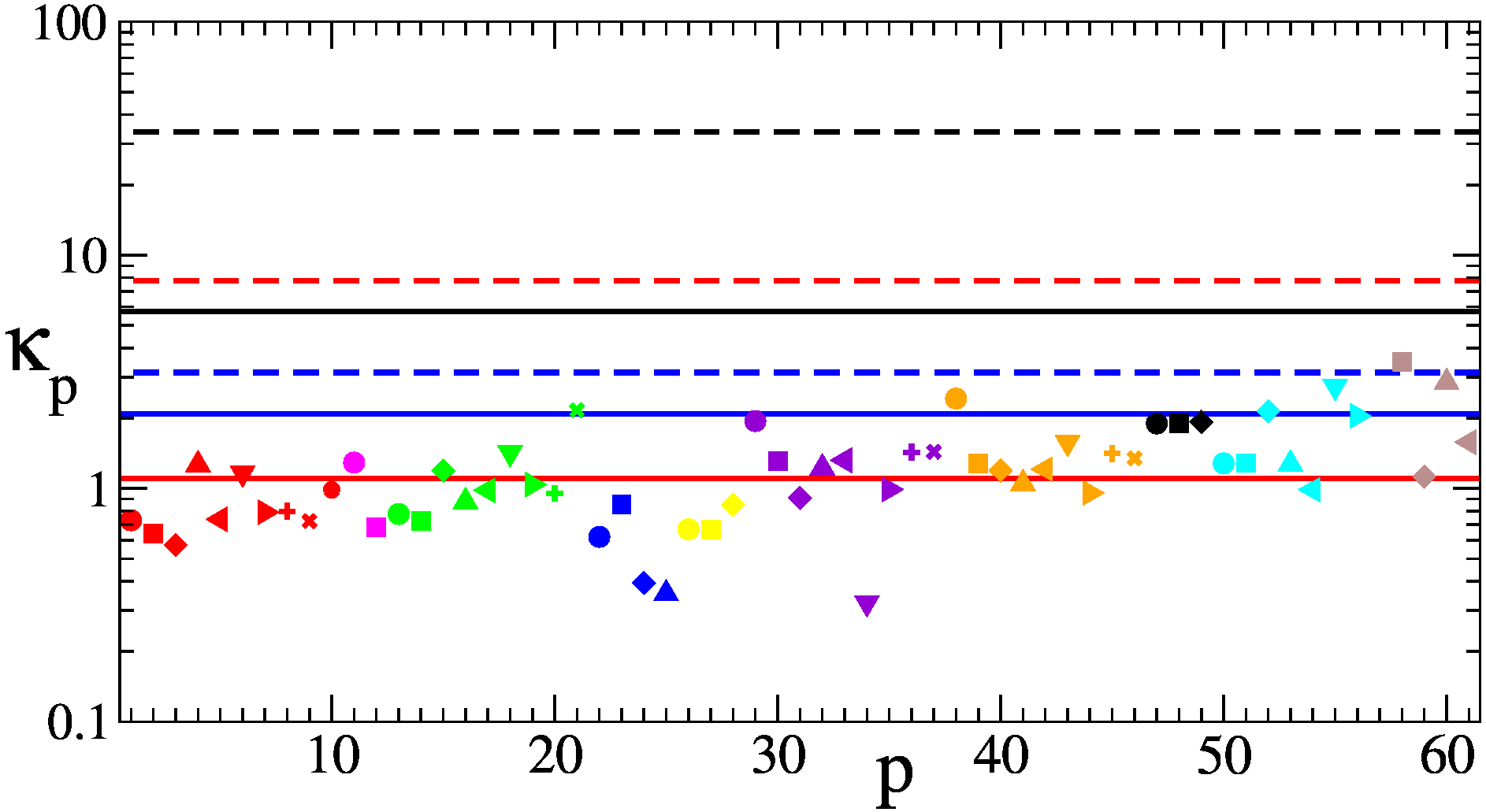} 
\end{center}
\vglue -0.1cm
\caption{
PageRank-CheiRank correlators $\kappa_p$ (\ref{eq11}) 
from the GPVM  are shown  as a function of the product 
index $p$ with the corresponding symbol from Fig.~\ref{fig5}.
PageRank-CheiRank  and 
ImportRank-ExportRank correlators are   
shown by solid and dashed lines respectively,
where the global correlator $\kappa$ (\ref{eq9})
is shown in black, the
correlator for countries $\kappa(c)$ (\ref{eq12}) is shown by red lines, 
the correlator for products $\kappa(p)$ (\ref{eq12})  is shown by blue lines. 
Here product number $p$ is counted in order of appearance in Table~\ref{table1}.
The data are given for year 2008 with $N_p=61, N_c=227, N=13847$.
}
\label{fig12}
\end{figure}

\begin{figure}[!ht]
\begin{center}
\includegraphics[width=0.47\textwidth]{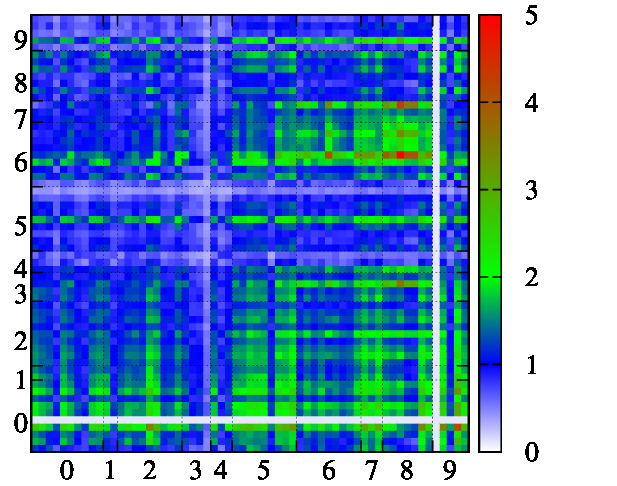} 
\end{center}
\vglue -0.1cm
\caption{
Product PageRank-CheiRank correlation matrix 
$\kappa_{p,p^\prime}$ (\ref{eq10}) for year 2008 with correlator values shown by color.
The code indexes $p$ and $p^{\prime}$ 
of all $N_p=61$ products are shown on $x$ and $y$ 
axes by their corresponding first digit (see Table~\ref{table1}).
}
\label{fig13}
\end{figure}

The values of correlators defined by Eqs.~(\ref{eq9})-(\ref{eq12})
are shown in Figs.~\ref{fig12},\ref{fig13} for a typical year 2008. 
For the global PageRank-CheiRank correlator we find $\kappa \approx 5.7$
(\ref{eq9})  while for Import-Export probabilities the
corresponding value is significantly larger with $\kappa \approx 33.7$.
Thus the trade volume ranking with its symmetry in products
gives an artificial increase of $\kappa$ by a significant factor.
A similar enhancement factor of Import-Export 
remains for correlators
in products $\kappa(p)$ and countries $\kappa(c)$ from Eq.~(\ref{eq12})
while for PageRank-CheiRank we obtain a moderate correlator values
around unity (see Fig.~\ref{fig12}). The PageRank-CheiRank correlator 
$\kappa_p$ (\ref{eq11}) for specific products have relatively low values
with $\kappa_p < 1$ for practically all products with $p \leq 45$ 
(we remind that here $p$ counts the products in the order of their appearance in the
Table~\ref{table1}, it is different from COMTRADE code number).

The correlation matrix of products $\kappa_{p p^{\prime}}$ (\ref{eq10}) 
is shown in Fig.~\ref{fig13}. This matrix is asymmetric and demonstrates the
existence of  relatively high correlations between
products {\it 73 Transport equipment}, 
{\it 65 Textile yarn, fabrics, made up articles, etc.}
and {\it 83 Travel goods, handbags and similar articles} that all are related with
transportation of products.

\subsection{Spectrum and eigenstates of WTN Google matrix}

\begin{figure}[!ht]
\begin{center}
\includegraphics[width=0.47\textwidth]{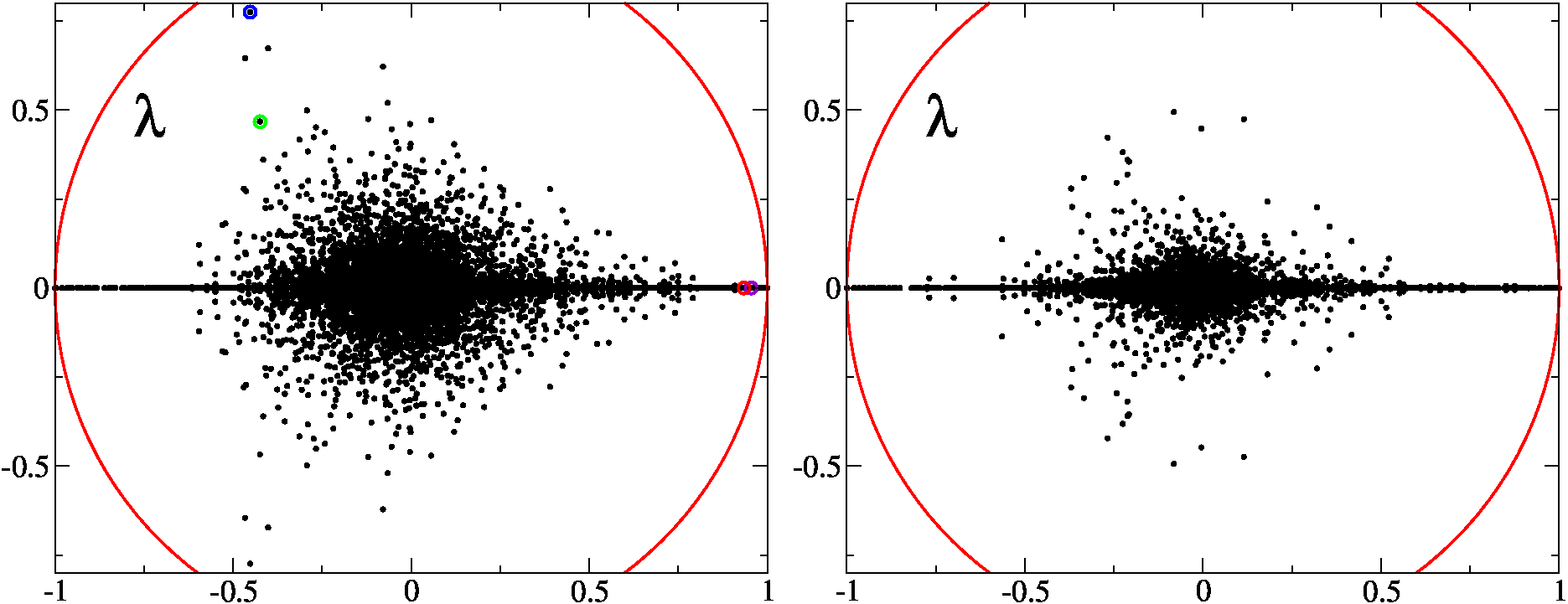} 
\end{center}
\vglue -0.1cm
\caption{
Spectrum of Google matrices $G$ (left panel) 
and $G^*$ (right panel) represented in the complex plane of $\lambda$.
The data are for year 2008 with $\alpha=1$, 
and  $N=13847$, $N_c=227$, $N_p=61$. Four eigenvalues marked by
colored circles are used for illustration of eigenstates
in Figs.~\ref{fig15},\ref{fig16}.
}
\label{fig14}
\end{figure}

\begin{figure}[!ht]
\begin{center}
\includegraphics[width=0.47\textwidth]{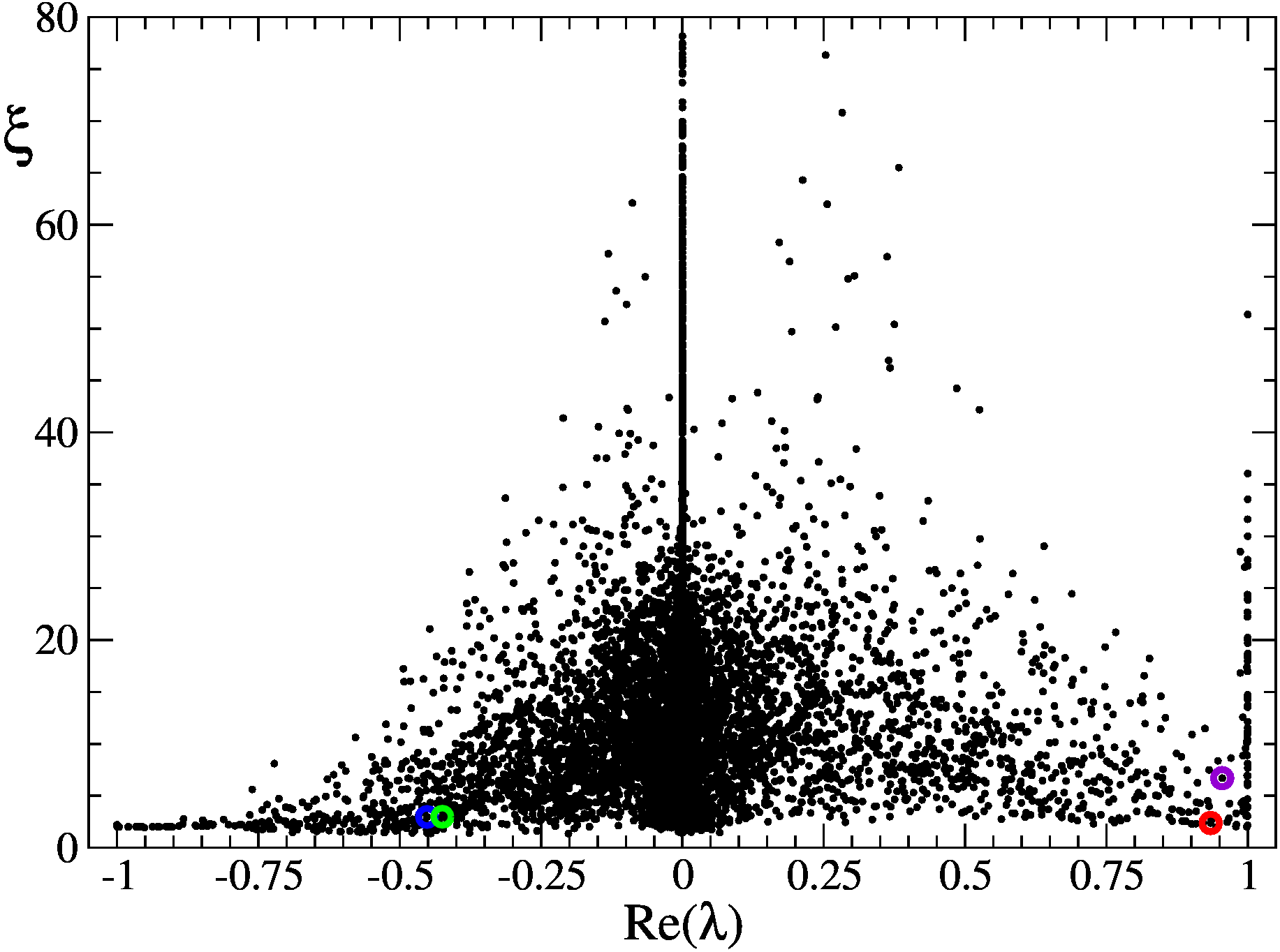} 
\end{center}
\vglue -0.1cm
\caption{
Inverse participation ratio (IPR) $\xi$ of all eigenstates of $G$ 
as a function of 
the real part of the corresponding eigenvalue $\lambda$
from the spectrum of Fig.~\ref{fig14}.
The eigenvalues marked by color circles are 
those from Fig.~\ref{fig14}
}
\label{fig15}
\end{figure}

Above we analyzed the properties of eigenstates of $G$ and $G^*$
at the largest eigenvalue $\lambda=1$. However, in total
there are $N$ eigenvalues and eigenstates. The results obtained
for the Wikipedia network \cite{wikispectrum} demonstrated that eigenstates
with large modulus of $\lambda$ correspond to certain specific communities
of the network. Thus it is interesting to study the spectral properties
of $G$ for the multiproduct WTN. The spectra of $G$ and $G^*$ 
are shown in Fig.~\ref{fig14} for year 2008. It is interesting to note
that for $G$ the spectrum shows some similarities with those of Wikipedia
(see Fig.1 in \cite{wikispectrum}). At $\alpha=1$ there are 12 and 7 degenerate
eigenvalues $\lambda=1$ for $G$ and $G^*$ respectively.
Thus the spectral gap appears only for $\alpha < 1$.
The dependence of IPR $\xi$ of eigenstates of $G$ on $Re \lambda$
is shown in Fig.~\ref{fig15}. The results show that $\xi \ll N$
so that the eigenstates are well localized on a certain group on nodes.

\begin{figure}[!ht]
\begin{center}
\includegraphics[width=0.47\textwidth]{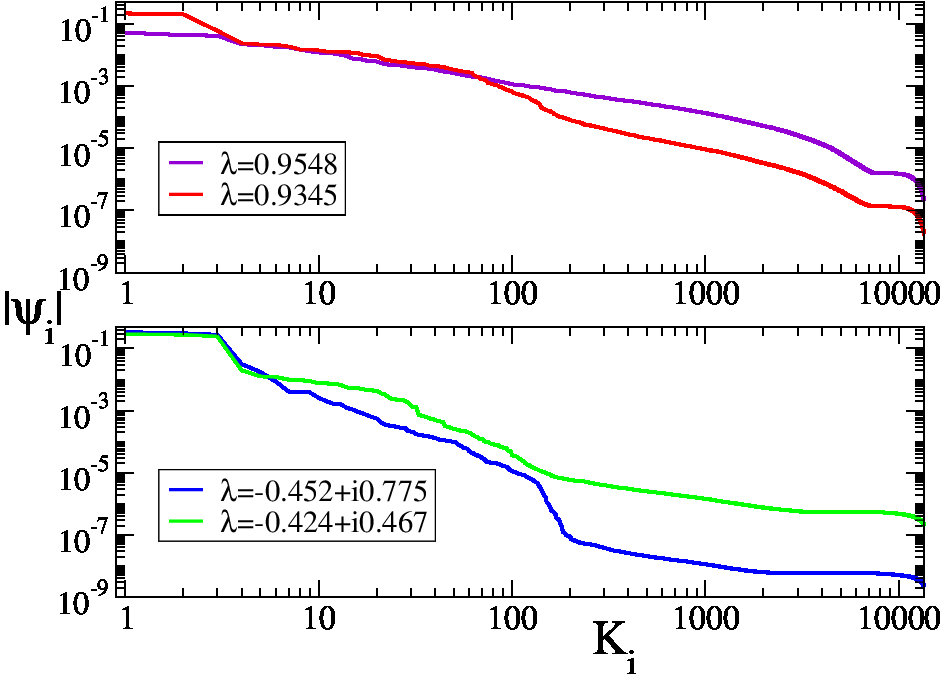} 
\end{center}
\vglue -0.1cm
\caption{
Eigenstate amplitudes $\vert\psi_i\vert$ ordered by 
its own decreasing amplitude order 
with index $K_i$ for 4 different eigenvalues of Fig.~\ref{fig14}
(states are normalized as $\sum_i\vert\psi_i\vert =1$). 
Top panel shows two example of real eigenvalues 
with $\lambda=0.9548$ and $\lambda=0.9345$
while bottom panel shows two eigenvalues with large imaginary part 
with $\lambda=0.452+i0.775$ and $\lambda=0.424+i0.467$. 
Node names (country, product) for top ten largest
amplitudes of these eigenvectors are shown in Table~\ref{table4}.
}
\label{fig16}
\end{figure}

The eigenstates $\psi_i$ can be ordered by their decreasing 
amplitude $|\psi_i|$ giving the eigenstate index $K_i$ with
the largest amplitude at $K_i=1$. The examples of four eigenstates
are shown in Fig.~\ref{fig16}. We see that 
the amplitude is mainly localized on a few 
top nodes in agreement of small values of $\xi \sim 4$
shown in Fig.~\ref{fig15}. The top ten amplitudes of these four eigenstates are 
shown in Table~\ref{table4} with corresponding names of countries and products.
We see that for a given eigenstate
these top ten nodes correspond to one product
clearly indicating strong links of trade
between certain countries. Thus for {\it 06 Sugar} we see strong link between
geographically close Mali and Guinea with further links to USA, Germany etc.
In a similar way for {\it 56 Fertilizers} there is a groups of Latin American countries
Brazil, Bolivia, Paraguay linked to Argentina, Uruguay etc.
We see a similar situation for products 57 and 52. These results confirm the observation 
established in \cite{wikispectrum} for Wikipedia
that the eigenstates with large modulus of $\lambda$ 
select interesting specific network communities.
We think that it would be interesting to investigate the properties of eigenstates
in further studies.

\subsection{Sensitivity to price variations}

Above we established the global mathematical structure of
multiproduct WTN and presented results on 
its ranking and spectral properties.
Such ranking properties bring new interesting
and important information about the WTN.
However, from the view point of economy
it is more important to analyze the effects of crisis
contamination and price variations. Such an analysis 
represents a complex task to which we hope to return
in our further investigations.
However, the knowledge of the global WTN structure
is an essential building block of this task
and we think that the presented results
demonstrate that this block is 
available now. 

\begin{figure}[!ht]
\begin{center}
\includegraphics[width=0.47\textwidth]{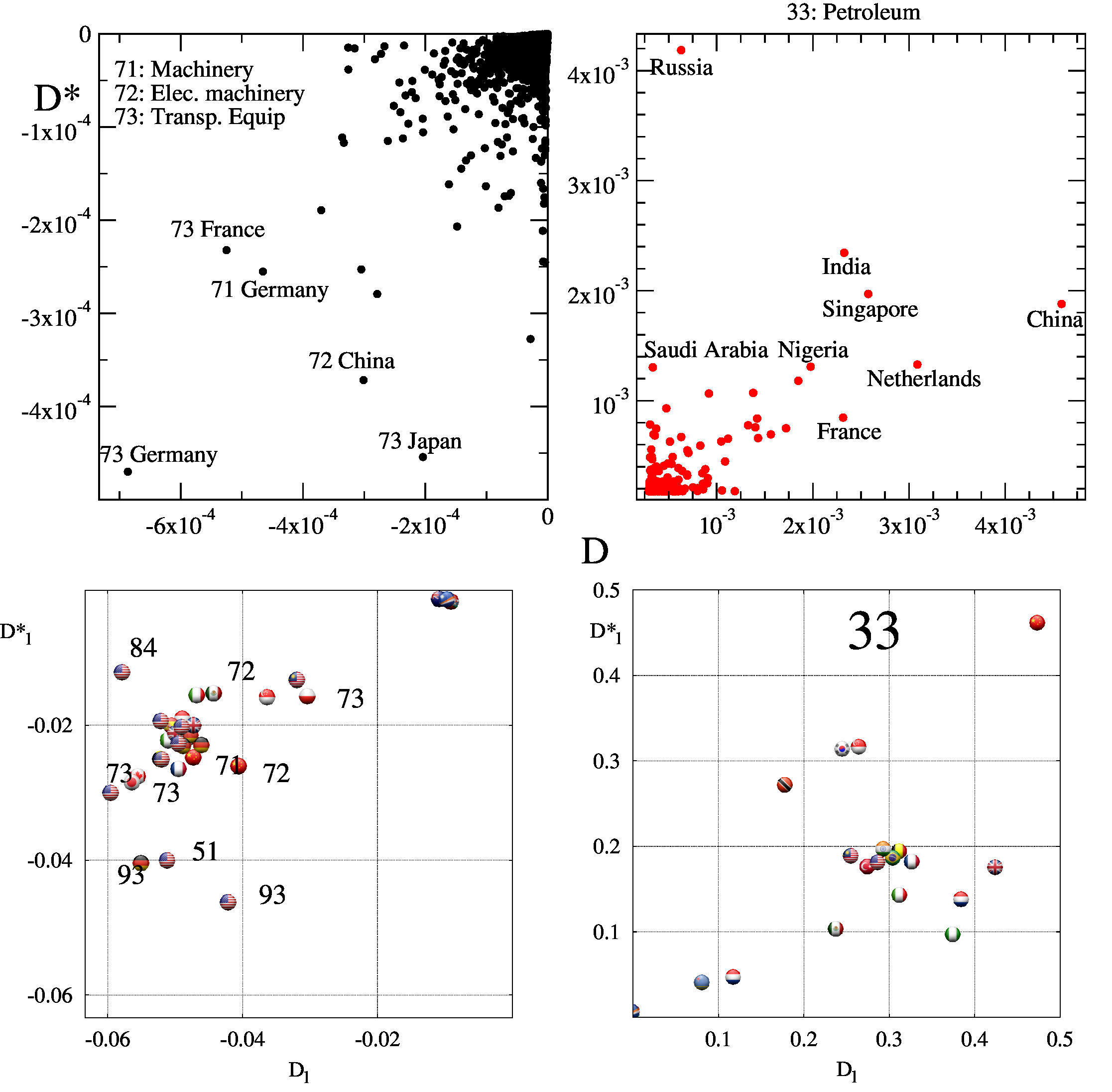} 
\end{center}
\vglue -0.1cm
\caption{
Derivatives $D=dP/d\delta_{33}$ and $D^*=dP^*/d\delta_{33}$ 
for a price variation $\delta_{33}$ of {\it 33 Petroleum and petroleum products"}
for year 2008.
Top left and right panels show the cases of negative and positive $D$ and $D^*$ 
respectively, with some products and countries labeled by their 2 digit code.
Bottom panels show the positive and negative cases
 of the logarithmic derivatives $D_l=D/P$ and $D^*_l=D^*/P^*$
for countries and products with $K_2 \leq 50$, 
where the flags and 2 digit codes for countries and products are shown
(in right panels only product 33 is present).
Codes are described in Table~\ref{table1}.
}
\label{fig17}
\end{figure}

Using the knowledge of WTN structure, we illustrate here
that it allows to obtain nontrivial results on 
sensitivity to price variations for certain products.
We consider as an example year 2008 and assume that the 
price of product {\it 33 Petroleum and petroleum products}
is increased by a relative fraction $\delta$
going from its unit value $1$ to $1+\delta$ (or $\delta=\delta_{33}$).
Then we compute the derivatives of probabilities
of PageRank $D=dP/d\delta = \Delta P/\delta$ and CheiRank 
$D^*=dP^*/d\delta= \Delta P^*/\delta$.
The computation is done for values of $\delta=0.01,0.03,0.05$
ensuring that the result is not sensitive to a
specific $\delta$ value. We also compute the logarithmic derivatives
$D_l= d \ln P/d\delta$, $D^*_l = d \ln P^*/d\delta$
which give us a relative changes of $P$, $P^*$.

\begin{figure}[!ht]
\begin{center}
\includegraphics[width=0.47\textwidth]{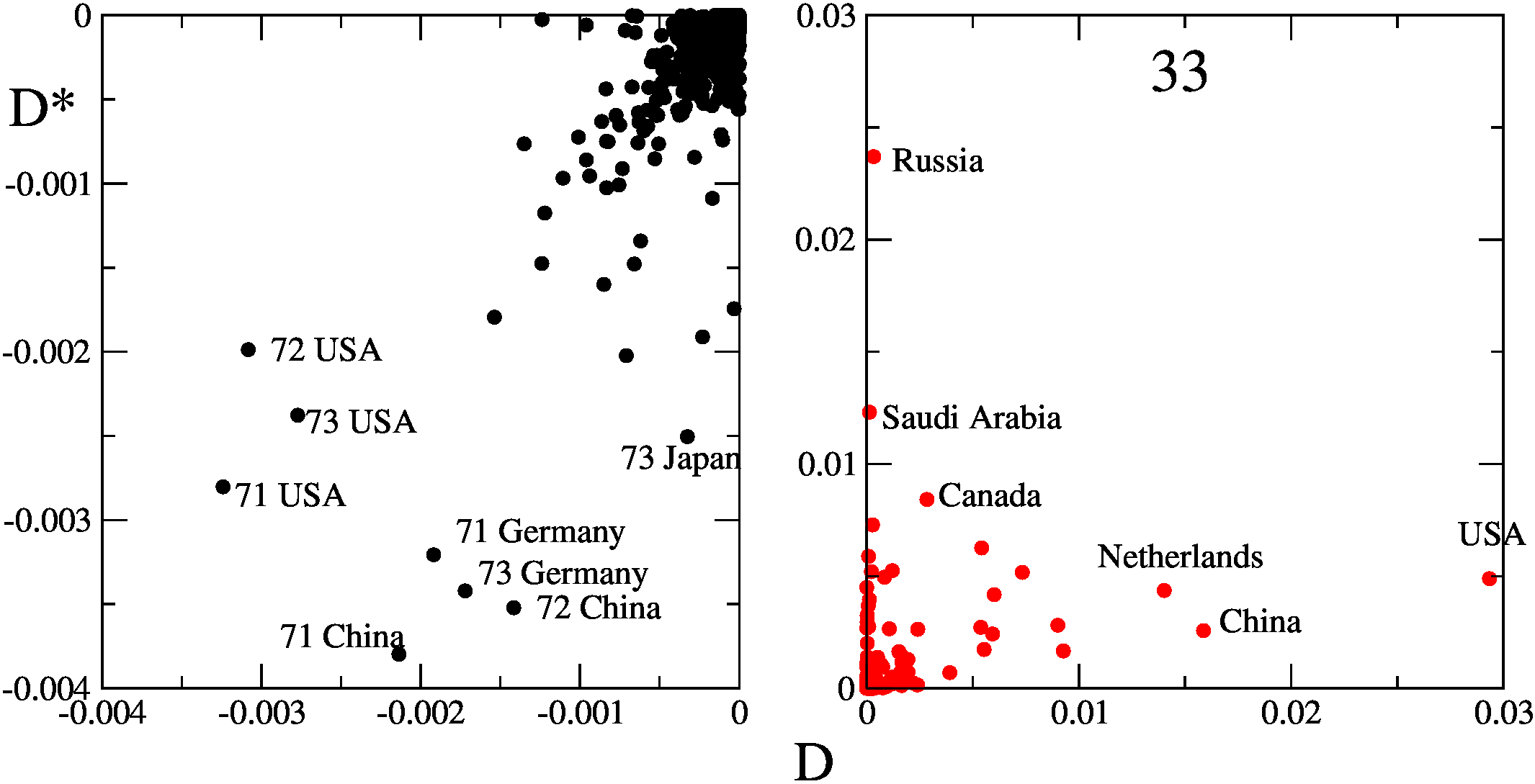} 
\end{center}
\vglue -0.1cm
\caption{
Same as in top panels of Fig.~\ref{fig17} but using probabilities
from the trade volume (\ref{eq3}).
}
\label{fig18}
\end{figure}

The results for the price variation $\delta_{33}$ 
of  {\it 33 Petroleum and petroleum products}
are shown in Fig.~\ref{fig17}. The derivatives for all WTN nodes
are shown on the planes $(D,D^*)$ and  $(D_l,D^*_l)$.
For $(D,D^*)$ the nodes are distributed in two sectors with
$D>0, D^*>0$ and $D<0, D^*<0$. The largest values with $D>0, D^*>0$
correspond to nodes of countries of product $33$ 
which are rich in petroleum (e.g. Russia, Saudi Arabia, Nigeria)
or those which have strong trade transfer of petroleum to other countries
(Singapore, India, China etc). It is rather natural that with the growth of petroleum
prices the rank probabilities $P,P^*$ of these countries grow.
A more unexpected effect is observed in the sector 
 $D<0, D^*<0$. Here we see that an increase of petroleum price
leads to a decrease of probabilities of 
nodes of countries Germany, France, China, Japan trading in 
machinery products $71, 72, 73$. 

For comparison we also compute the
derivatives $D, D^*$, $D_l, D^*_l$ from the probabilities 
(\ref{eq3})
defined by the trade volume of Import-Export instead of PageRank-CheiRank.
The results are shown in Fig.~\ref{fig18} for petroleum price variation
to be compared with Fig.~\ref{fig17}.
The distribution of $D,D^*$ is rather different from those
values obtained with PageRank-CheiRank probabilities.
This is related to the fact that PageRank and CheiRank
take into account the global network structure while the trade volume
gives only local relations in trade links between countries.
The difference between these two methods becomes even more
striking for logarithmic derivatives $D_l, D^*_l$.
Indeed, for the trade volume ranking the variation of 
probabilities $P^*, P$ due to price variation of a given product
can be computed analytically taking into account
the trade volume change with $\delta_p$. The computations give
$D_{cp}=(1-f_p)P_{cp}$, 
${D^*}_{cp}=(1-f_p){P^*}_{cp}$ 
for a derivative of of probability of product $p$ and country $c$ over the price of product
$\delta_p$ and $D_{cp^{\prime}}=-f_p P_{cp^{\prime}}$,  
$D_{cp^{\prime}}=-f_p P_{cp^{\prime}}$,  ${D^*}_{cp^{\prime}}=-f_p {P^*}_{cp^{\prime}}$ (if $p^{\prime} \neq p$),
where $f_p$ is a fraction of product $p$ in the world trade.
From these expressions we see that the logarithmic derivatives
are independent of country and product. Indeed, for
the case of Fig.~\ref{fig18} we obtain analytically and by direct numerical computations that
$D_l =D^*_l=-0.2022$ (for all countries if $p^{\prime} \neq p=33$)
and $D_l =D^*_l=0.7916$ (for all countries if $p^{\prime} = p=33$).
Due to simplicity of this case we do not show it in Fig.~\ref{fig18}.

\begin{figure}[!ht]
\begin{center}
\includegraphics[width=0.47\textwidth]{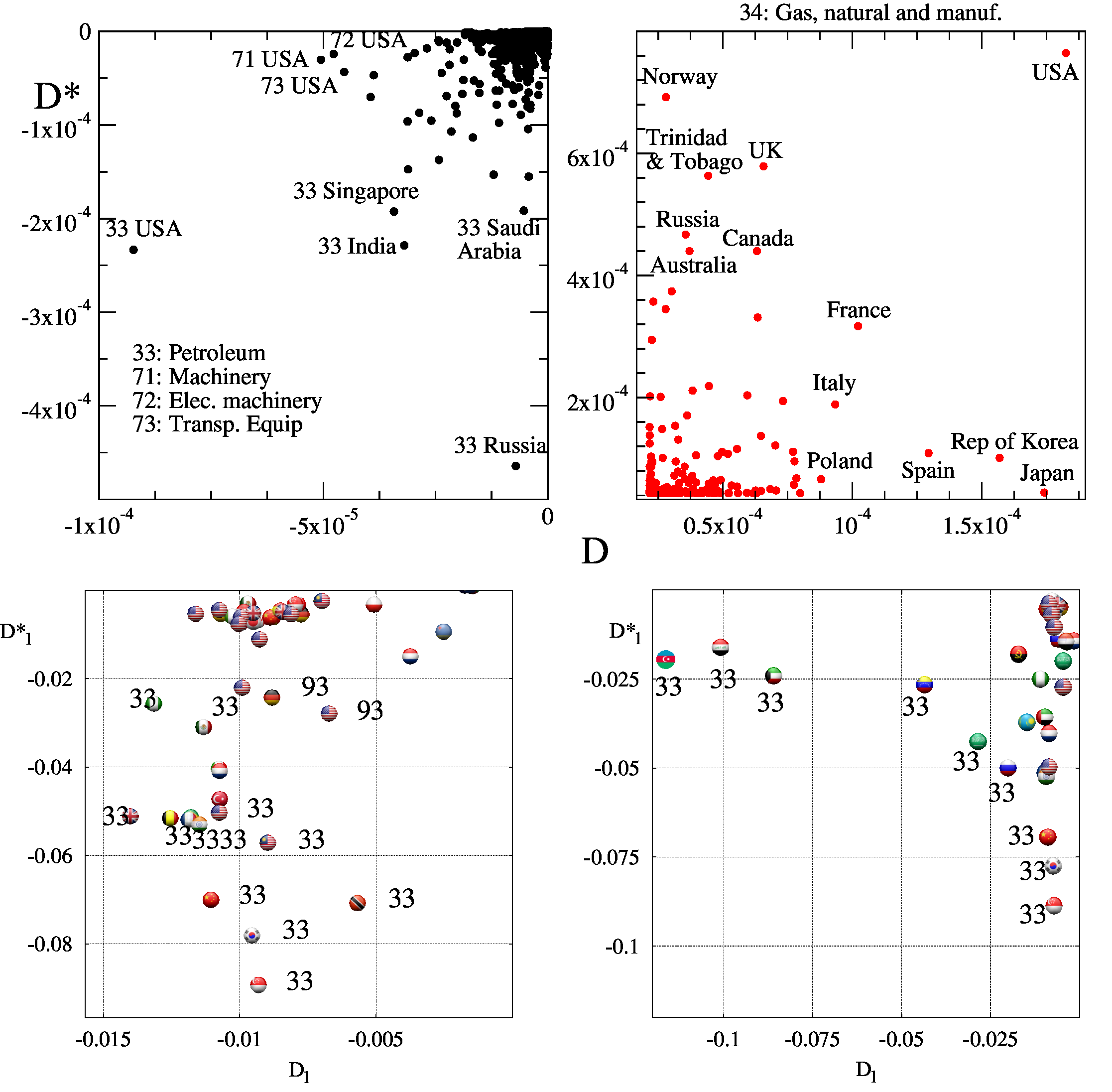} 
\end{center}
\vglue -0.1cm
\caption{
Derivative of $P$ and $P^*$ ($D$ and $D^*$ respectively)
for a price variation of {\it 34 Gas, natural and manufactured}
for 2008.
Top left and right panels show the cases of negative and positive 
sectors of $D$ and $D^*$ 
respectively, with some 
products and countries labeled by their 2 digit code and names
(in top right panel all points correspond to product 34).
Bottom panels show the cases of the logarithmic derivatives $D_l$ and $D^*_l$
for countries and products with 
$K_2 \leq 50$ (bottom left panel) and $K,K^* \leq 25$ 
(bottom right panel);
flags and 2 digit codes for countries and products are shown. 
In bottom right panel ($K,K^*\leq25$) we do not show the case of Sudan 
({\it 73 Transport equipment}) which has values of 
$(D_l,D^*_l)=(2\times10^{-4},1.75\times10^{-2})$. 
Codes are described in Table~\ref{table1}.
}
\label{fig19}
\end{figure}

The results for price variation of
{\it 34 Gas, natural and manufactured}
are presented in Fig.~\ref{fig19}
showing derivatives of PageRank and CheiRank
probabilities over $\delta_{34}$.
We see that for absolute derivatives $D,D^*$ 
the mostly affected are now nodes of 
gas producing countries for the sector $D,D^*>0$,
while for the sector   $D,D^* < 0$ the mostly affected are
countries linked to petroleum production or trade,
plus USA with products 71,72,73. For the sector
of logarithmic derivatives
$D_l, D^*_l <0$ among top 
$K_2$ and $K,K^*$ nodes we find nodes of countries 
of product  33 and also 93. 

Thus the analysis of derivatives provides an interesting new information
of sensitivity of world trade to price variations.

\subsection{World map of CheiRank-PageRank trade balance}

On the basis of the obtained WTN Google matrix we can now analyze the trade balance
in various products between the world countries.
Usually economists consider the export and import of a given country 
as it is shown in Fig.~\ref{fig1}.  Then the trade balance of a given country $c$
can be defined making summation over all products:  
\begin{equation}
B_c=\sum_p (P^*_{cp} - P_{cp})/\sum_p (P^*_{cp} + P_{cp}) = (P^*_{c} - P_{c})/(P^*_{c} + P_{c}) .
\label{eq13} 
\end{equation}
In economy, $P_c, P^*_c$ are defined via the probabilities of trade volume
$\hat{P}_{cp}, \hat{P}^*_{cp}$  from (\ref{eq3}). In our approach,
we define $P_{cp}, P^*_{cp}$ as PageRank and CheiRank probabilities. 
In contrast to the trade volume our approach takes into account the 
multiple network links between nodes. 

\begin{figure}[!ht]
\begin{center}
\includegraphics[width=0.48\textwidth]{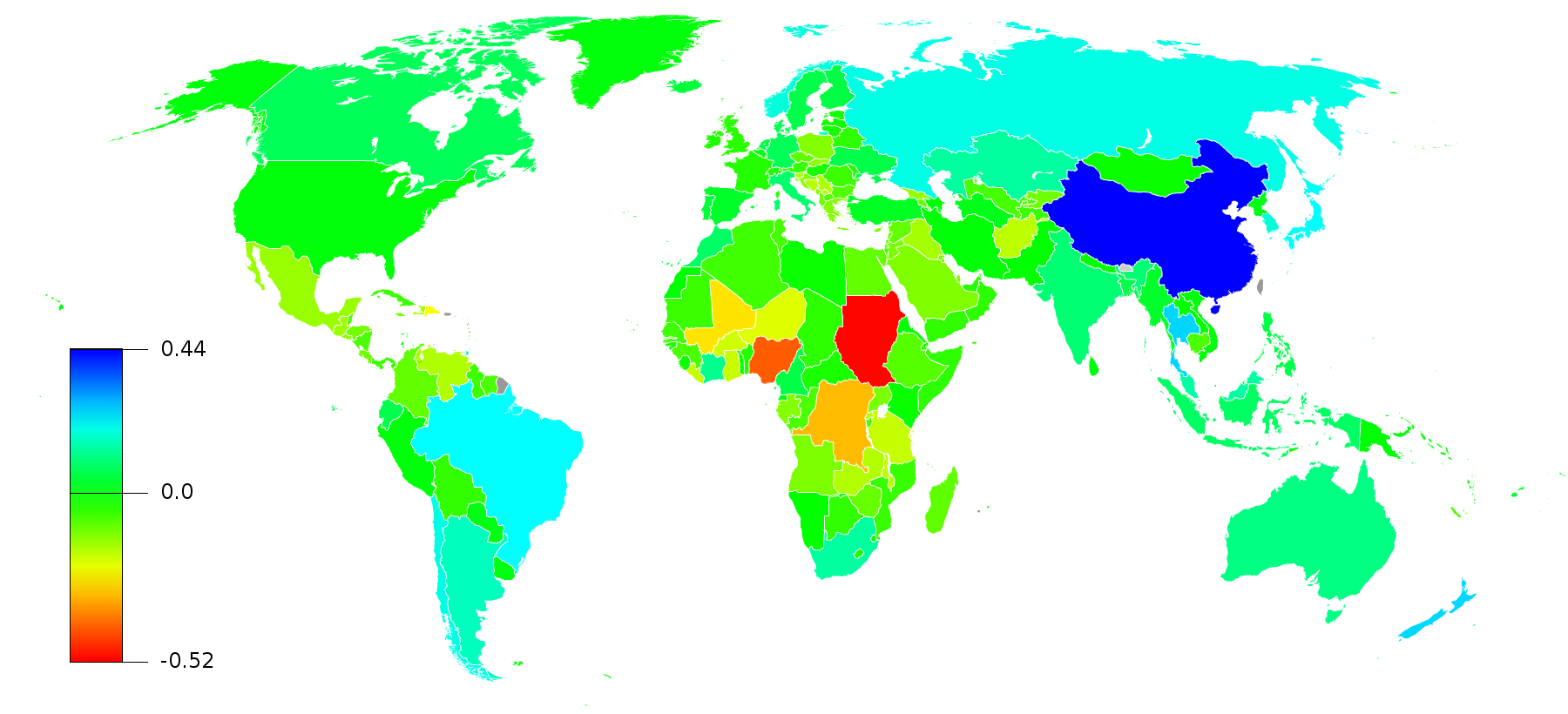}\\
\includegraphics[width=0.48\textwidth]{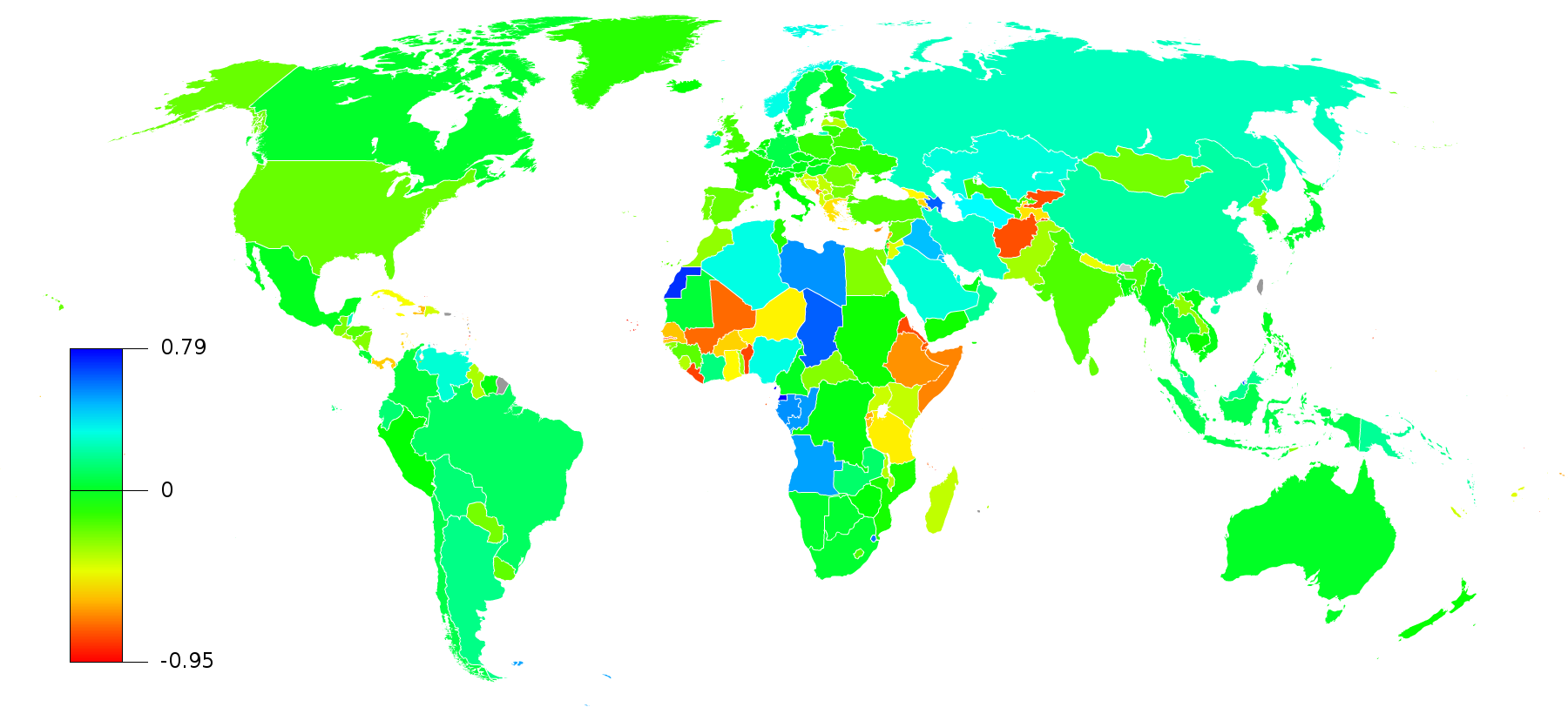}
\end{center}
\vglue -0.1cm
\caption{
World map of probabilities balance $B_c=(P^*_c-P_c)/(P^*_c+P_c)$
determined for each of $N_c=227$ countries in year 2008.
Top panel: probabilities $P^*_c, P_c$ are given by 
CheiRank and PageRank vectors; bottom panel:
probabilities are computed from the trade volume 
of Export-Import (\ref{eq3}). Names of countries can be find at
\cite{worldmap}.
}
\label{fig20}
\end{figure}

 The comparison of the 
world trade balance obtained by these two methods is 
shown in Fig.~\ref{fig20}. We see that the leadership of China
becomes very well visible in CheiRank-PageRank balance map
while it is much less pronounced in the trade volume balance.
The Google matrix analysis also highlights the dis-balance of trade network
of Nigeria (strongly oriented on petroleum export and  machinery import)
and Sudan. It is interesting to note that the positive
CheiRank-PageRank balance is mainly located in the countries of BRICS
(Brazil, Russia, India, China, South Africa).
In contrast to that, the usual trade volume balance
highlights Western Sahara and Afganistan at large positive and
negative trade balance in 2008.

\begin{figure}[!ht]
\begin{center}
\includegraphics[width=0.48\textwidth]{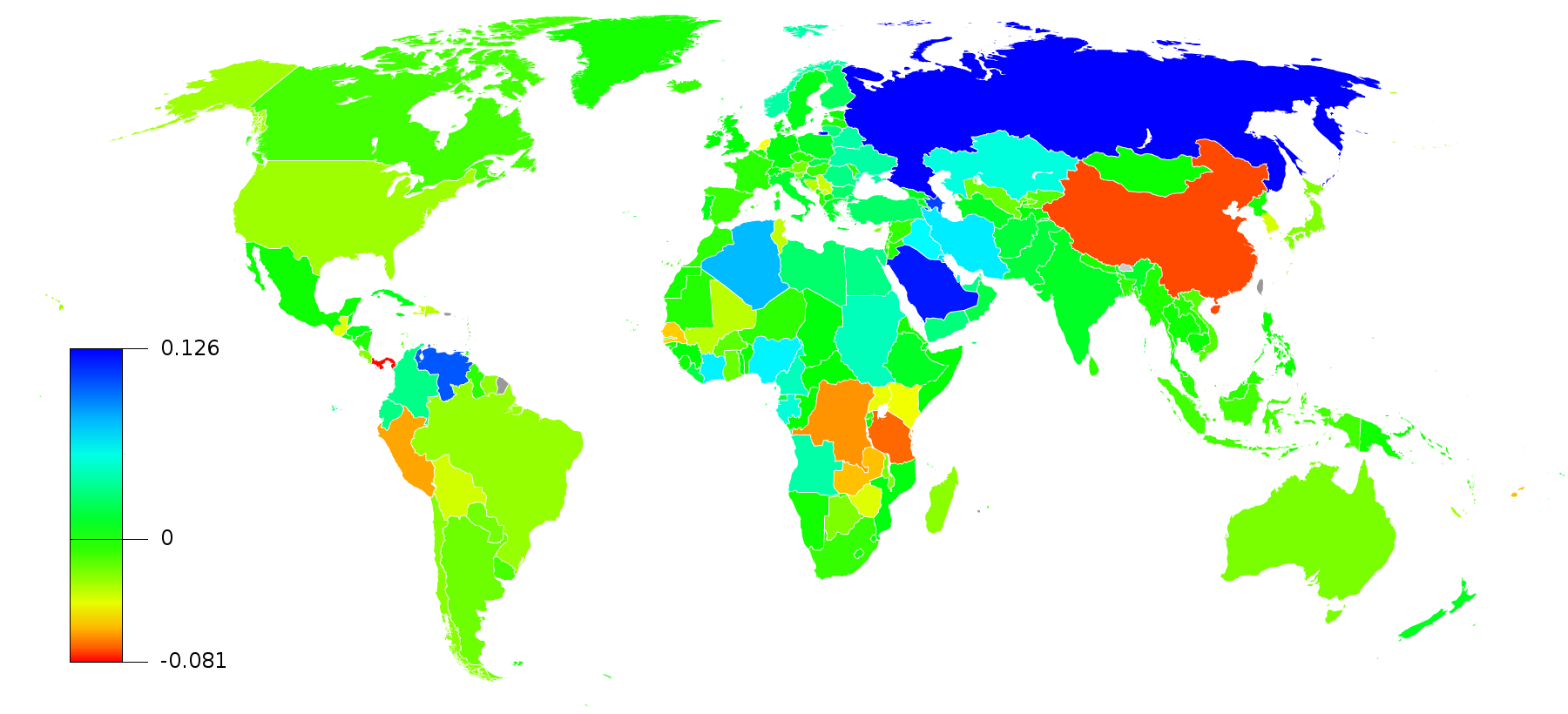}\\
\includegraphics[width=0.48\textwidth]{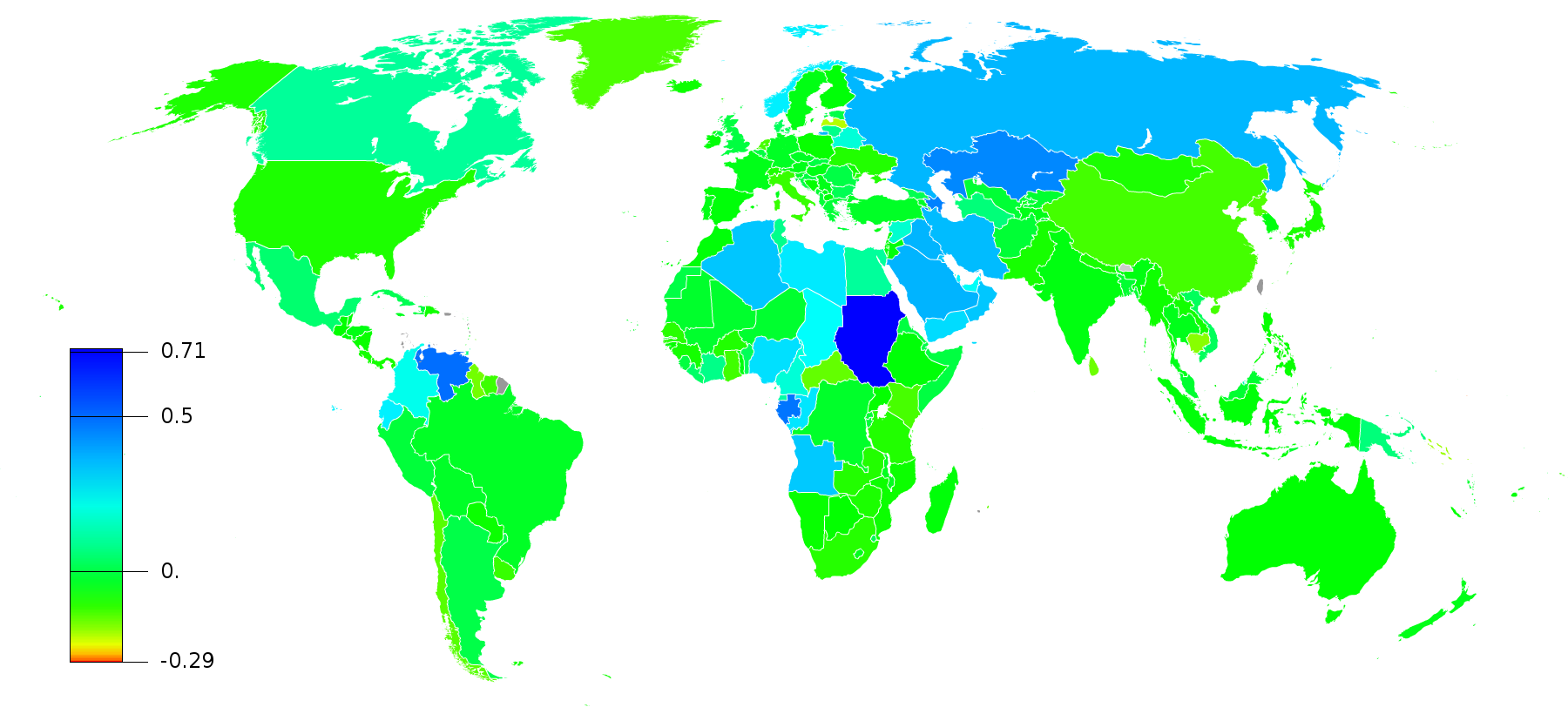}
\end{center}
\vglue -0.1cm
\caption{
Derivative of probabilities balance $dB_c/d \delta_{33}$
over petroleum price $\delta_{33}$ for  year 2008.
Top panel: balance of countries $B_c$  is determined from 
CheiRank and PageRank vectors as in the top panel of Fig.~\ref{fig20}; 
bottom panel: $B_c$ values are 
computed from the trade volume  as in the bottom panel of Fig.~\ref{fig20}.
Names of countries can be find at \cite{worldmap}.
}
\label{fig21}
\end{figure}

We can also determine the sensitivity of trade balance to
price variation of a certain product $p$ computing the balance derivative
$dB_c/d\delta_p$. The world map sensitivity in respect 
to price of petroleum $p=33$
is shown in Fig.~\ref{fig21} for the above two methods
of definition of probabilities $P_c, P^*_c$ in (\ref{eq13}).
For the CheiRank-PageRank balance we see that the derivative $dB_c/d\delta_{33}$
is positive for countries producing  petroleum 
(Russia, Saudi Arabia,  Venezuela)
while the highest negative derivative appears for China
which economy is happened to be very sensitive to petroleum price.
The results from the trade volume computation of $dB_c/d\delta_p$,
shown in Fig.~\ref{fig21}, give 
rather different distribution of derivatives over
countries with maximum for Sudan and minimum 
for the Republic of Nauru (this country has very small area
and is not visible in the bottom panel of Fig.~\ref{fig21}), while for
China the balance
looks to be not very sensitive 
to $\delta_{33}$
(in contrast to the CheiRank-PageRank method).
This happens due to absence of links between nodes
in the trade volume computations while
the CheiRank-PageRank approach 
takes links into account and recover hidden 
trade relations between products and countries.

\begin{figure}[!ht]
\begin{center}
\includegraphics[width=0.48\textwidth]{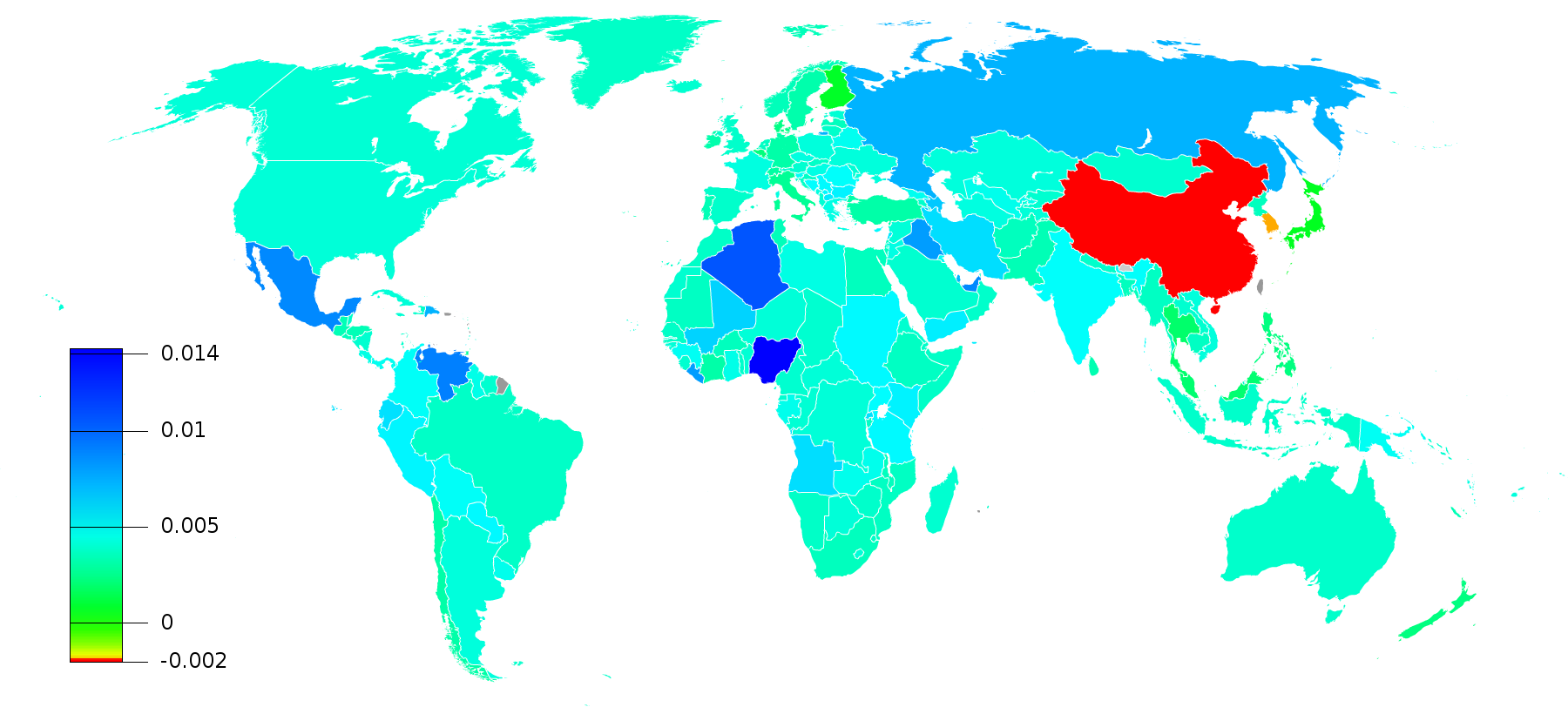}\\
\includegraphics[width=0.48\textwidth]{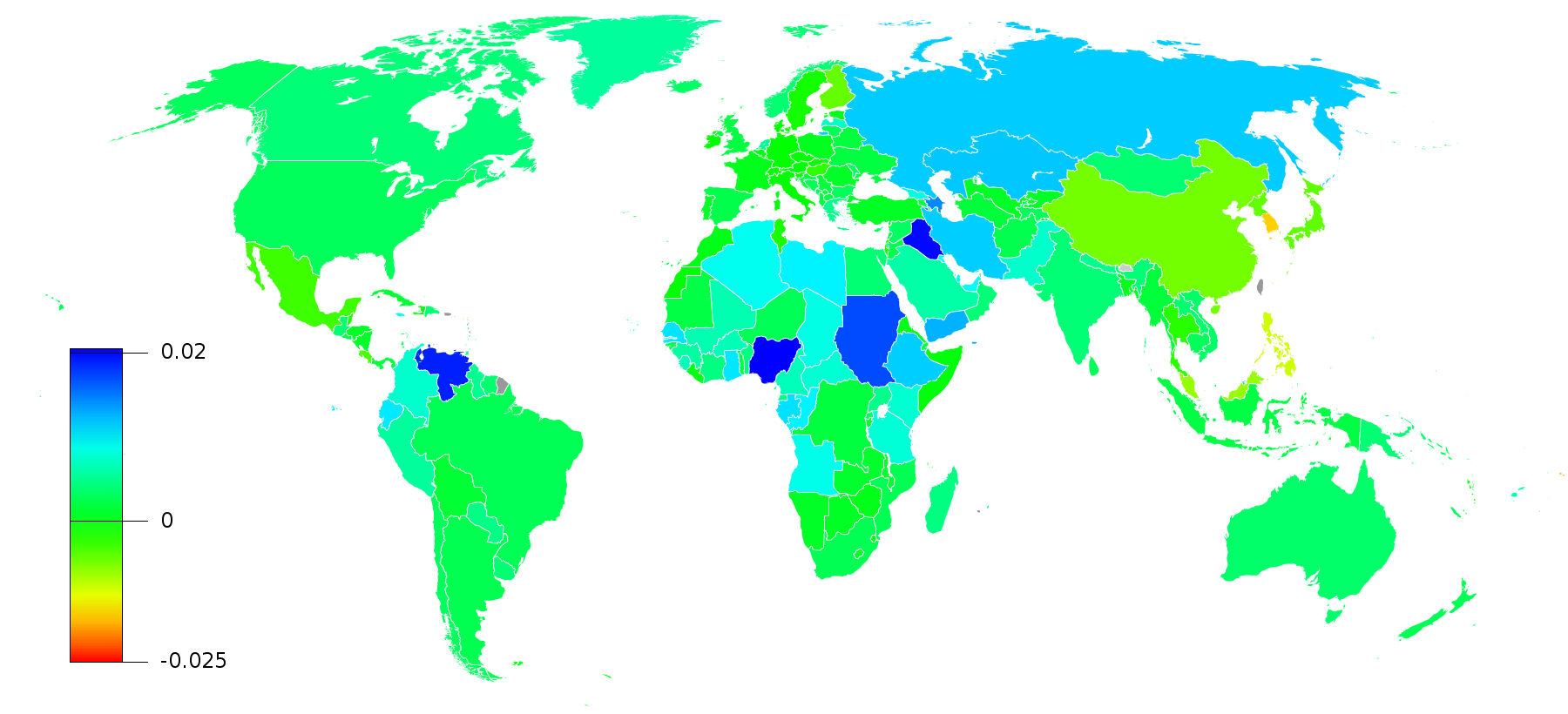}
\end{center}
\vglue -0.1cm
\caption{
Derivative of partial probability balance of 
product $p$ defined as $dB_{cp}/d \delta_{33}$
over petroleum price $\delta_{33}$ for  year 2008; 
here $B_{cp}=(P^*_{cp}-P_{cp})/(P^*_c+P_c)$
and $p=72$ ({\it 72 Electrical machinery ...} from Table~\ref{table1});
the product balance of countries $B_{cp}$  is determined from 
CheiRank and PageRank vectors (top panel)
and from the trade volume  of Export-Import  (\ref{eq3})
(bottom panel).
Names of countries can be find at \cite{worldmap}.
}
\label{fig22}
\end{figure}

This absence of links in the trade volume approach
becomes also evident if we consider the derivative of
the partial trade balance for a given product $p$ defined as
\begin{equation}
B_{cp}=(P^*_{cp} - P_{cp})/\sum_p (P^*_{cp} + P_{cp}) = 
(P^*_{cp} - P_{cp})/(P^*_{c} + P_{c}) ,
\label{eq14} 
\end{equation}
so that the global country balance is $B_c=\sum_p B_{cp}$.
Then the sensitivity of partial balance of a given product $p$
in respect to a price variation of a product $p^{\prime}$
is given by the derivative
$d B_{cp}/ d \delta_{p^{\prime}}$. The sensitivity
for balance of product $p=72$ ({\it 72 Electrical machinery ...})
in respect to petroleum $p^{\prime}=33$ price  variation
$\delta_{33}$ is shown for the CheiRank-PageRank balance 
in Fig.~\ref{fig22} (top panel)
indicating sensitivity of trade balance
of product $p=72$ at the petroleum $p^{\prime}=33$ price variation.
We see that China has a negative derivative for this partial balance.
In contrast, the computations based on
the trade volume (Fig.~\ref{fig22} bottom panel)
give a rather different distribution of derivatives
$d B_{cp}/ d \delta_{p^{\prime}}$ over countries.
In the trade volume approach the derivative $d B_{cp}/ d \delta_{p^{\prime}}$
appears due to the renormalization of total trade volume
and nonlinearity coming from the ratio of probabilities.
We argue that the CheiRank-PageRank approach 
treats the trade relations between products and 
countries on a significantly
more advanced level taking into account all the complexity
of links in the multiproduct world trade.

\begin{figure}[!h]
\begin{center}
\includegraphics[width=0.40\textwidth]{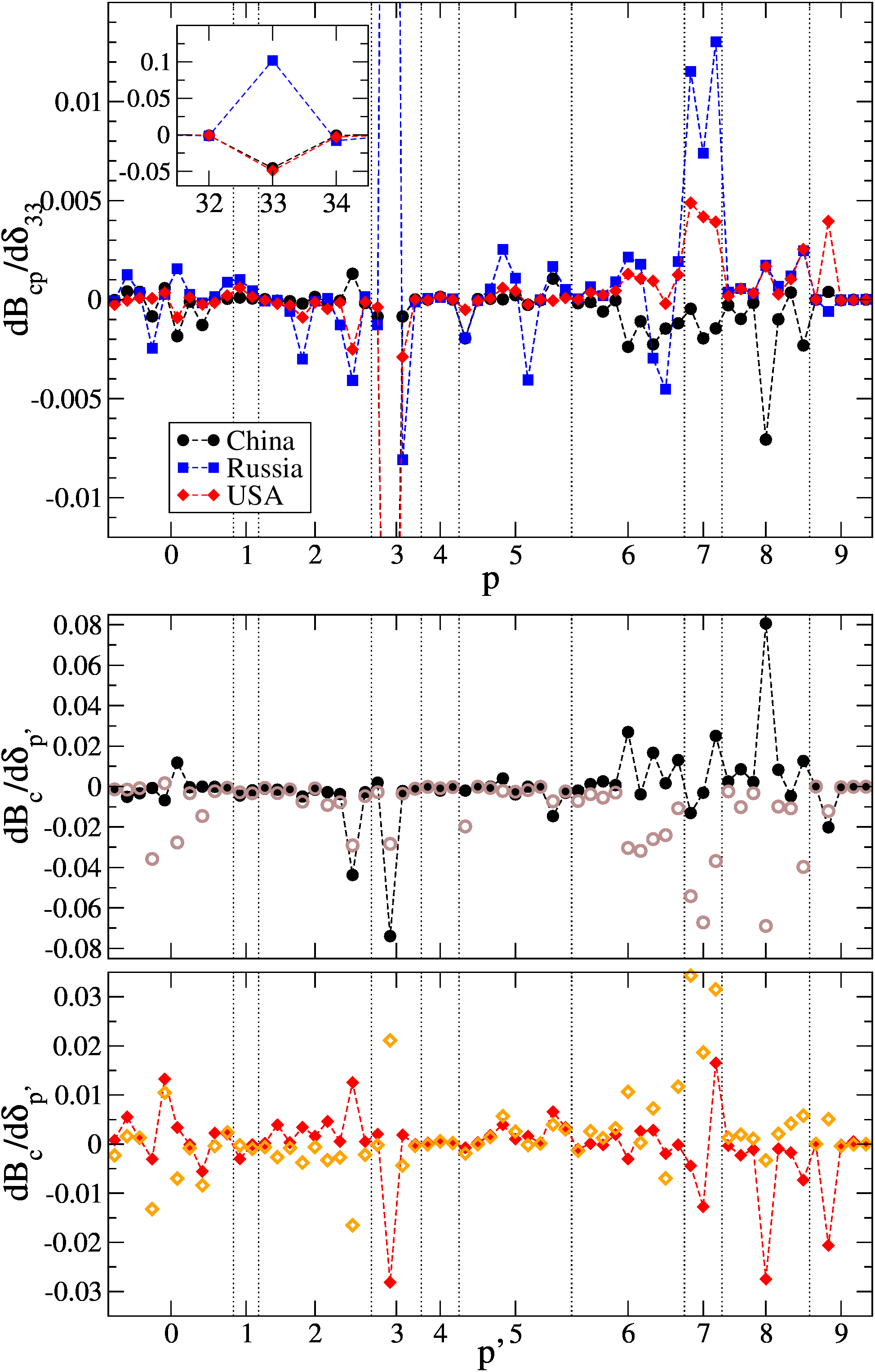} 
\end{center}
\vglue -0.1cm
\caption{ 
Top panel:
derivative $dB_{cp}/d \delta_{33}$ 
of partial probability balance $B_{cp}$ of product $p$
over petroleum price $\delta_{33}$ for year 2008 and countries: 
China (black circles), Russia (blue squares) and USA (red diamonds); 
inset panel shows the products of digit $3$ including the diagonal
term $p=33$ being out of scale in the main panel;
here $B_{cp}=(P^*_{cp}-P_{cp})/(P^*_c+P_c)$ (\ref{eq14}).
Center (China) and bottom (USA) panels show
derivative  $dB_{c}/d \delta_{p^{\prime}}$ 
of country total probability balance $B_c$
over price $ \delta_{p^{\prime}}$ of product $p^{\prime}$ for year 2008;
derivatives of balance without  diagonal term 
($dB_{c}/d \delta_{p^{\prime}}-dB_{cp^{\prime}}/d \delta_{p^{\prime}}$) 
are represented by open  circles and open diamonds 
for China and USA respectively.
The product balance of countries $B_{cp}$  and $B_c$ are determined from 
CheiRank and PageRank vectors. The vertical dotted lines mark
the first digit of product index $p$ or $p^{\prime}$ from Table~\ref{table1}.
}
\label{fig23}
\end{figure}

Using the CheiRank-PageRank approach we determine the sensitivity
of partial balance of all $61$ products in respect to petroleum price
variation $\delta_{33}$ for China, Russia and USA, as shown in Fig.~\ref{fig23}
(top panel). We see that the diagonal derivative $d B_{c33}/d \delta_{33}$
is positive for Russia but is negative for China and USA.
Even if USA produce petroleum its sensitivity is negative
due to a significant import of petroleum to USA.
For non-diagonal derivatives over $\delta_{33}$
we find positive sensitivity of Russia and USA for products $p=71,72,73$
while for China it is negative.   Other product partial balances sensitive
to petroleum are e.g. {\it 84 Clothing} for China for which 
expensive petroleum gives an increase of transportation costs; 
negative derivative of balance in metal products $p=67,68$ for Russia
due to fuel price increase; positive derivative 
for  {\it  93 Special transact. ...} of USA.

The sensitivity of country balance $B_c$ to price variation $\delta_{p^\prime}$ for
all products is shown in Fig.~\ref{fig23} for China (middle panel)
and USA (bottom panel). We find that the balance of China is very sensitive
to $p^\prime=33,84$ and indeed, these products play an important role in its
economy with negative and positive derivatives respectively.
For USA the trade balance is also very sensitive to these  
two products $p^\prime=33,84$ but the derivative is negative in both cases. 
We also present the derivative of balance without diagonal term
($d (B_c -B_{cp^\prime})/d \delta_{p^\prime}$) for China and USA.
This quantity shows that for USA all other 
products give a positive derivative for $p^\prime=33$ but the 
contribution of petroleum import gives the global negative derivative
of the total USA balance. In a similar way for China for 
$p^\prime=84$ all products, except the diagonal one $p^\prime=84$, give a negative
sensitivity for balance but the diagonal contribution
of  $p^\prime=84$ gives the final positive derivative
of China total balance in respect to $\delta_{84}$.

\begin{figure}[!h]
\begin{center}
\includegraphics[width=0.40\textwidth]{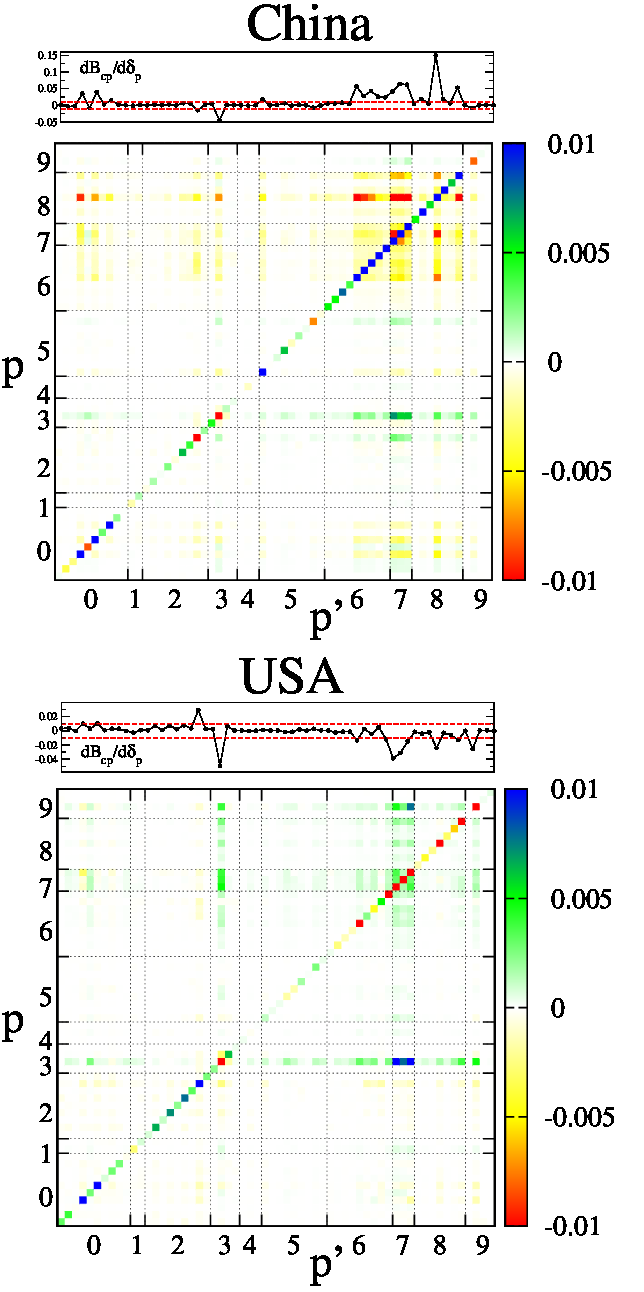} 
\end{center}
\vglue -0.1cm
\caption{
China (top) and USA (bottom) examples of derivative $dB_{cp}/d  \delta_{p^{\prime}}$ 
of partial probability balance $B_{cp}$ of product $p$ 
over price $ \delta_{p^{\prime}}$ of product 
$p^{\prime}$ for year 2008 .
Diagonal terms, given by $dB_{cp}/d  \delta_{p}$ vs. $p=p^{\prime}$,
are shown on the top panels of each example.
Products $p^{\prime}$ and $p$ are shown in $x$-axis and $y$-axis respectively
(indexed as in Table~\ref{table1}),
while $dB_{cp}/d  \delta_{p^{\prime}}$ is represented by colors with a threshold 
value given by $-0.01$ and $0.01$ for negative and positive values respectively,
also shown in red dashed lines on top panels with diagonal terms.
Dotted lines mark the first digit of Table~\ref{table1}.
Here $B_{cp}$ are defined by CheiRank and PageRank probabilities.
}
\label{fig24}
\end{figure}

The CheiRank-PageRank approach allows to determine cross-product
sensitivity of partial trade balance
computing the derivative  $dB_{cp}/d  \delta_{p^{\prime}}$ 
shown in Fig.~\ref{fig24} for China and USA. The derivatives are
very different for two countries showing a structural difference of their 
economies. Thus for China the cross-derivative (at $p \neq p^{\prime}$)
are mainly negative (except a few lines around $p=33$)
but the diagonal terms  $dB_{cp}/d  \delta_p$ are mainly positive.
In contrast, for USA the situation is almost the opposite.
We attribute this to the leading role of China in export
and the leading role of USA in import.
However, a detailed analysis of these cross-products derivatives and 
correlations require further more detailed analysis.
We think that the presented cross-product sensitivity 
plays and important role in the multiproduct
trade network that are highlighted by the Google matrix analysis
developed here. This analysis allows to 
determine efficiently the sensitivity of multiproduct trade  
in respect to price variations of various products.

\section{Discussion}

In this work we have developed the Google matrix analysis of the multiproduct 
world trade network.
Our approach allows to treat all world countries on equal democratic grounds
independently of their richness keeping the contributions of trade products
proportional to their fractions in the world trade. As a result of this 
approach we have obtained a reliable ranking of world countries and products
for years 1962 - 2010. The Google analysis captures the years with crises
and also shows that after averaging over all 
world countries some products are export oriented
while others are import oriented. This feature is absent in
the usual Import-Export analysis based on trade volume
which gives a symmetric orientation of products after 
such an averaging.

\begin{figure}[!h]
\begin{center}
\includegraphics[width=0.40\textwidth]{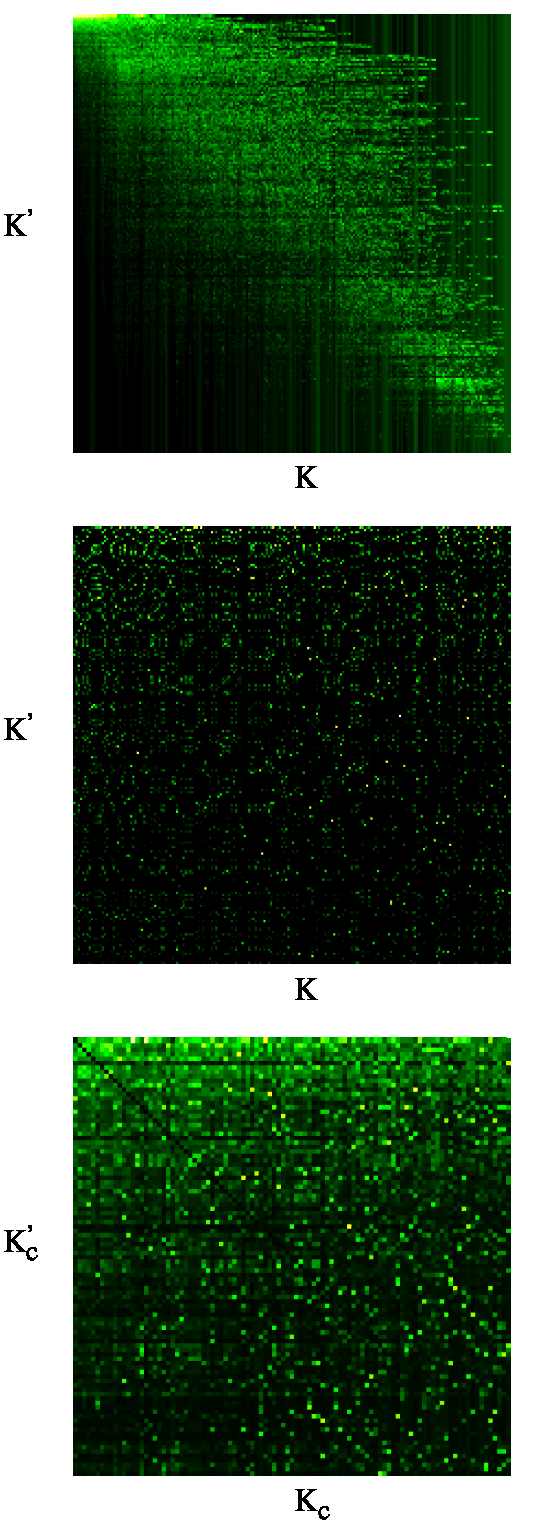} 
\end{center}
\vglue -0.1cm
\caption{ Google matrix $G_{KK^{\prime}}$ representation for $2008$ 
with $\alpha=0.5$ ordered by PageRank index $K$value 
(where $K=K^\prime=1$ is on top left corner).
Top panel shows the whole Google matrix ($N=N_c \times N_p = 227 \times 61 =13847$) 
with coarse-graining of $N \times N$ elements down to $200\times200$ shown cells.
Center panel represents the top corner of 
the full Google matrix with $K,K^\prime\leq200$.
Bottom panel shows the coarse-grained Google matrix 
for  countries 
for the top 100 countries ($K_c,K_c^\prime \leq 100$).
Color changes from black at minimal matrix element
to white at maximal element, $\alpha=0.5$.
}
\label{fig25}
\end{figure}

The WTN matrix analysis determines the trade balance 
for each country not only in trade volume but also in CheiRank-PageRank
probabilities which take into account multiple trade links between
countries which are absent in the usual Export-Import considerations.
The CheiRank-PageRank balance highlights in a clear manner the leading WTN role
of new rising economies of China and other BRICS countries.
This analysis also allows to determine the sensitivity
of trade network to price variations of various products that opens
new possibilities for analysis of cross-product price influence
via network links absent in the standard Export-Import analysis. 

We think that this work makes only first steps in the 
development of WTN matrix analysis of multiproduct world trade.
Indeed, the global properties of the Google matrix of 
multiproduct WTN should be studied in more detail
since the statistical properties of matrix elements of $G$,
shown in Fig.~\ref{fig23} for year 2008, are still
not well understood (e.g. visible patterns present in the coarse-grained 
representation of $G$ in Fig.~\ref{fig23}).

Even if the UN COMTRADE database contains a lot of information there 
are still open questions if all essential economic aspects are 
completely captured in this database. Indeed, the COMTRADE data
for trade exchange are diagonal in products since there are
no interactions (trade) between products. However,
this feature may be a weak point of collected data
since in a real economy there is a transformation of
some products into some other products
(e.g. metal and plastic are transferred to cars and machinery).
It is possible that additional data should be collected to
take into account the existing interactions between products.
There are also some other aspects of services and various
other activities which are not present in the COMTRADE database
and which can affect the world economy.

One of the important missing element of COMTRADE 
are financial flows between countries.
Indeed, the product {\it 93 Special trans. ...}
(see Tables~\ref{table1},\ref{table2}) partially
takes into account the financial flows but it is clear that
the interbank flows are not completely reported in the database.
In fact the Wold Bank Web (WBW) really exists 
(e.g. a private person can transfer money from his bank account
to another person account using
SWIFT code) but the flows on the WBW remain completely hidden
and not available for scientific analysis. The size on interbank networks
are relatively small (e.g. the whole Federal Reserve of USA 
has only $N \approx 6600$ bank nodes \cite{soramaki} and 
there are only about $N \approx 2000$ bank nodes in Germany \cite{craig}).
Thus the WBW size of the whole world is about a few tens of thousands
of nodes and the Google matrix analysis should be well adapted 
for WBW. We consider that there are many similarities between the
multiproduct WTN and the WBW, where financial transfers are performed with
various financial products so that the above WTN analysis should be
well suited for the WBW. The network approach to the WBW
flows is now at the initial development stage (see e.g.
\cite{soramaki,craig,garratt}) but hopefully the 
security aspects will be handled in an efficient manner
opening possibilities for the Google matrix analysis of the WBW.
The joint analysis of trade and financial flows between
world countries would allow to reach a scientific understanding of 
peculiarities of such network flows and to control
in an efficient way financial and petroleum crises.

The developed Google matrix analysis 
of multiproduct world trade allows to
establish hidden dependencies between various 
products and countries and opens new prospects
for further studies of this interesting 
complex system of world importance.

\section{Acknowledgments}
We thank the representatives of UN COMTRADE \cite{comtrade}
for providing us with the friendly access to this database.
This research is supported in part by the EC FET Open project
``New tools and algorithms for directed network analysis''
(NADINE $No$ 288956). We thank Barbara Meller (Deutsche Bundesbank, Zentrale)
for constructive critical remarks.

\onecolumn

\begin{tiny}
\begin{table}[!ht]%
\caption {Codes and names of the 61 products from COMTRADE 
Standard International Trade Classification (SITC) Rev. 1.}
\centering %
\begin{tabular}{|cl|cl|} 
\hline
code	&	name	&	code	&	name\\
\hline
00	&	Live animals			&	54	&	Medicinal and pharmaceutical products	\\
01	&	Meat and meat preparations			&	55	&	Perfume materials, toilet \& cleansing preptions	\\
02	&	Dairy products and eggs			&	56	&	Fertilizers, manufactured	\\
03	&	Fish and fish preparations			&	57	&	Explosives and pyrotechnic products	\\
04	&	Cereals and cereal preparations			&	58	&	Plastic materials, etc.	\\
05	&	Fruit and vegetables			&	59	&	Chemical materials and products, nes	\\
06	&	Sugar, sugar preparations and honey			&	61	&	Leather, lthr. Manufs., nes \& dressed fur skins	\\
07	&	Coffee, tea, cocoa, spices \& manufacs. Thereof			&	62	&	Rubber manufactures, nes	\\
08	&	Feed. Stuff for animals excl. Unmilled cereals			&	63	&	Wood and cork manufactures excluding furniture	\\
09	&	Miscellaneous food preparations			&	64	&	Paper, paperboard and manufactures thereof	\\
11	&	Beverages			&	65	&	Textile yarn, fabrics, made up articles, etc.	\\
12	&	Tobacco and tobacco manufactures			&	66	&	Non metallic mineral manufactures, nes	\\
21	&	Hides, skins and fur skins, undressed			&	67	&	Iron and steel	\\
22	&	Oil seeds, oil nuts and oil kernels			&	68	&	Non ferrous metals	\\
23	&	Crude rubber including synthetic and reclaimed			&	69	&	Manufactures of metal, nes	\\
24	&	Wood, lumber and cork			&	71	&	Machinery, other than electric	\\
25	&	Pulp and paper			&	72	&	Electrical machinery, apparatus and appliances	\\
26	&	Textile fibres, not manufactured, and waste			&	73	&	Transport equipment	\\
27	&	Crude fertilizers and crude minerals, nes			&	81	&	Sanitary, plumbing, heating and lighting fixt.	\\
28	&	Metalliferous ores and metal scrap			&	82	&	Furniture	\\
29	&	Crude animal and vegetable materials, nes			&	83	&	Travel goods, handbags and similar articles	\\
32	&	Coal, coke and briquettes			&	84	&	Clothing	\\
33	&	Petroleum and petroleum products			&	85	&	Footwear	\\
34	&	Gas, natural and manufactured			&	86	&	Scientif \& control instrum, photogr gds, clocks	\\
35	&	Electric energy			&	89	&	Miscellaneous manufactured articles, nes	\\
41	&	Animal oils and fats			&	91	&	Postal packages not class. According to kind	\\
42	&	Fixed vegetable oils and fats			&	93	&	Special transact. Not class. According to kind	\\
43	&	Animal and vegetable oils and fats, processed			&	94	&	Animals, nes, incl. Zoo animals, dogs and cats	\\
51	&	Chemical elements and compounds			&	95	&	Firearms of war and ammunition therefor	\\
52	&	Crude chemicals from coal, petroleum and gas			&	96	&	Coin, other than gold coin, not legal tender	\\
53	&	Dyeing, tanning and colouring materials			&		&		\\
\hline
\end{tabular}
\label{tab1}\label{table1}
\end{table}
\end{tiny}

\begin{table}[ht]%
\caption {Columns represent data: codes of  61 products of COMTRADE SITC Rev.1, 
ImportRank and ExportRank $\hat{K}=\hat{K^*}$  in year 2008, 
product fraction in global trade volume in 2008,
$\hat{K}=\hat{K^*}$ in 1998, product fraction in 1998.}
\centering %
\begin{tabular}{|c|c|c|c|c||c|c|c|c|c|} 
\hline
code	& $\hat{K}$(08)	&$\%$ vol(08)	&$\hat{K}$(98)	&	$\%$ vol (98)	&code	& $\hat{K}$(08)	&$\%$ vol(08)	&$\hat{K}$(98)	&$\%$ vol (98)	\\
\hline
00	&	53	&	0.10	&	51	&	0.17	&	54	&	9	&	2.89	&	16	&	1.88	\\
01	&	27	&	0.69	&	26	&	0.83	&	55	&	25	&	0.76	&	28	&	0.79	\\
02	&	34	&	0.44	&	34	&	0.56	&	56	&	30	&	0.55	&	43	&	0.36	\\
03	&	28	&	0.63	&	22	&	0.99	&	57	&	58	&	0.03	&	57	&	0.04	\\
04	&	21	&	1.07	&	19	&	1.13	&	58	&	15	&	1.95	&	13	&	2.07	\\
05	&	19	&	1.16	&	18	&	1.50	&	59	&	22	&	1.04	&	20	&	1.13	\\
06	&	49	&	0.23	&	44	&	0.36	&	61	&	51	&	0.19	&	42	&	0.37	\\
07	&	33	&	0.47	&	29	&	0.73	&	62	&	26	&	0.73	&	24	&	0.85	\\
08	&	38	&	0.39	&	36	&	0.45	&	63	&	35	&	0.43	&	32	&	0.61	\\
09	&	40	&	0.34	&	41	&	0.39	&	64	&	20	&	1.14	&	17	&	1.79	\\
11	&	31	&	0.54	&	31	&	0.65	&	65	&	18	&	1.40	&	11	&	2.46	\\
12	&	47	&	0.24	&	35	&	0.49	&	66	&	17	&	1.71	&	14	&	2.01	\\
21	&	56	&	0.05	&	53	&	0.11	&	67	&	7	&	3.63	&	10	&	2.74	\\
22	&	37	&	0.39	&	48	&	0.32	&	68	&	11	&	2.27	&	15	&	1.95	\\
23	&	44	&	0.26	&	50	&	0.22	&	69	&	13	&	2.04	&	12	&	2.12	\\
24	&	39	&	0.35	&	30	&	0.65	&	71	&	2	&	11.82	&	1	&	15.03	\\
25	&	43	&	0.29	&	45	&	0.34	&	72	&	3	&	10.42	&	3	&	12.26	\\
26	&	50	&	0.22	&	37	&	0.45	&	73	&	4	&	10.06	&	2	&	12.38	\\
27	&	41	&	0.33	&	47	&	0.33	&	81	&	42	&	0.31	&	46	&	0.34	\\
28	&	16	&	1.92	&	25	&	0.84	&	82	&	23	&	0.93	&	21	&	1.03	\\
29	&	48	&	0.24	&	40	&	0.39	&	83	&	45	&	0.26	&	49	&	0.26	\\
32	&	24	&	0.82	&	39	&	0.42	&	84	&	10	&	2.42	&	6	&	3.44	\\
33	&	1	&	14.88	&	4	&	5.02	&	85	&	29	&	0.59	&	27	&	0.79	\\
34	&	14	&	2.04	&	23	&	0.99	&	86	&	12	&	2.25	&	8	&	2.95	\\
35	&	46	&	0.26	&	52	&	0.17	&	89	&	6	&	3.72	&	5	&	4.54	\\
41	&	57	&	0.03	&	58	&	0.04	&	91	&	61	&	0.00	&	61	&	0.00	\\
42	&	32	&	0.49	&	38	&	0.44	&	93	&	5	&	3.92	&	9	&	2.92	\\
43	&	54	&	0.08	&	55	&	0.08	&	94	&	59	&	0.01	&	59	&	0.01	\\
51	&	8	&	3.01	&	7	&	3.07	&	95	&	55	&	0.08	&	54	&	0.11	\\
52	&	52	&	0.11	&	56	&	0.05	&	96	&	60	&	0.00	&	60	&	0.00	\\
53	&	36	&	0.39	&	33	&	0.60	&		&		&		&		&		\\
\hline
\end{tabular}
\label{tab2}\label{table2}
\end{table}

\begin{table}[ht]%
\caption {Top 20 ranks for global PageRank $K$, CheiRank $K^*$, 
2dRank $K_2$, ImportRank $\hat{K}$ and ExportRank $\hat{K}^*$ 
for given country and product code for year 2008.
}
\centering %
\begin{tabular}{|l|lc|lc|lc|lc|lc|} 
\hline
\#& 
\multicolumn{2}{c|}{$K$}&
\multicolumn{2}{c|}{$K^*$}&
\multicolumn{2}{c|}{$K_2$}&
\multicolumn{2}{c|}{$\hat{K}$}&
\multicolumn{2}{c|}{$\hat{K}^*$}\\
& 
\multicolumn{2}{c|}{country \& code}&
\multicolumn{2}{c|}{country \& code}&
\multicolumn{2}{c|}{country \& code}&
\multicolumn{2}{c|}{country \& code}&
\multicolumn{2}{c|}{country \& code}\\
\hline
1	&	USA	&	33	&	Russia	&	33	&	Germany	&	73	&	USA	&	33	&	China	&	72	\\
2	&	USA	&	73	&	China	&	84	&	USA	&	73	&	USA	&	71	&	Russia	&	33	\\
3	&	USA	&	71	&	Germany	&	73	&	USA	&	33	&	USA	&	72	&	China	&	71	\\
4	&	USA	&	93	&	Japan	&	73	&	USA	&	71	&	USA	&	73	&	Germany	&	73	\\
5	&	Germany	&	73	&	USA	&	73	&	India	&	33	&	Japan	&	33	&	Germany	&	71	\\
6	&	USA	&	72	&	China	&	72	&	Singapore	&	33	&	China	&	72	&	Saudi Arabia	&	33	\\
7	&	France	&	73	&	USA	&	33	&	Germany	&	71	&	China	&	33	&	USA	&	71	\\
8	&	Germany	&	71	&	India	&	33	&	USA	&	72	&	Germany	&	71	&	Japan	&	73	\\
9	&	Singapore	&	33	&	USA	&	71	&	France	&	73	&	Germany	&	73	&	USA	&	73	\\
10	&	India	&	33	&	China	&	71	&	Netherlands	&	33	&	Netherlands	&	33	&	Japan	&	71	\\
11	&	China	&	33	&	Singapore	&	33	&	USA	&	93	&	Germany	&	72	&	USA	&	72	\\
12	&	Netherlands	&	33	&	Saudi Arabia	&	33	&	Nigeria	&	33	&	China	&	71	&	China	&	89	\\
13	&	France	&	33	&	Germany	&	71	&	Germany	&	72	&	USA	&	89	&	Germany	&	72	\\
14	&	UK	&	71	&	USA	&	72	&	China	&	72	&	Italy	&	33	&	China	&	84	\\
15	&	UK	&	73	&	France	&	73	&	China	&	71	&	Germany	&	33	&	Japan	&	72	\\
16	&	Germany	&	72	&	Thailand	&	3	&	UK	&	33	&	South Korea	&	33	&	South Korea	&	72	\\
17	&	USA	&	89	&	Kazakhstan	&	33	&	Germany	&	93	&	France	&	73	&	France	&	73	\\
18	&	South Korea	&	33	&	U. Arab Emir.	&	33	&	China	&	33	&	China	&	28	&	Italy	&	71	\\
19	&	France	&	71	&	USA	&	28	&	South Korea	&	33	&	Germany	&	93	&	U. Arab Emir.	&	33	\\
20	&	Sudan	&	73	&	Netherlands	&	33	&	Australia	&	33	&	India	&	33	&	Germany	&	93	\\
\hline
\end{tabular}
\label{tab3}\label{table3}
\end{table}

\begin{table}[ht]%
\caption {Top 10 values of 4 different eigenvectors from Fig.~\ref{fig16}.
The corresponding eigenvalues form left to right are
  $\lambda=0.9548$, $\lambda=0.9345$, $\lambda=0.452+i0.775$ and 
$\lambda=0.424+i0.467$. 
There is only one product in each of these top 10 list nodes which are:
57 \emph{Explosives and pyrotechnic products};
06 \emph{Sugar, sugar preparations and honey};
56 \emph{Fertilizers, manufactured};
52 \emph{Crude chemicals from coal, petroleum and gas}.
}
\centering %
\begin{tabular}{|c||c|l||c|l||c|l||c|l|} 
\hline
$K_i$	&	$\vert\psi_i\vert$	&	country	&	$\vert\psi_i\vert$	&	country	&	$\vert\psi_i\vert$	&	country	&	$\vert\psi_i\vert$	&	country	\\
\hline
	&		&	prod: 57	&		&	prod:06	&		&	prod:56	&		&	prod:52	\\
\hline
1	&	0.052	&	USA	&	0.216	&	Mali	&	0.332	&	Brazil	&	0.288	&	Japan	\\
2	&	0.044	&	Tajikistan	&	0.201	&	Guinea	&	0.304	&	Bolivia	&	0.279	&	Rep. of Korea	\\
3	&	0.042	&	Kyrgyzstan	&	0.059	&	USA	&	0.274	&	Paraguay	&	0.245	&	China	\\
4	&	0.022	&	France	&	0.023	&	Germany	&	0.031	&	Argentina	&	0.020	&	Australia	\\
5	&	0.021	&	Mexico	&	0.021	&	Mexico	&	0.017	&	Uruguay	&	0.013	&	USA	\\
6	&	0.018	&	Italy	&	0.021	&	Canada	&	0.009	&	Chile	&	0.012	&	U Arab Em	\\
7	&	0.018	&	Canada	&	0.018	&	UK	&	0.004	&	Portugal	&	0.010	&	Canada	\\
8	&	0.015	&	Germany	&	0.015	&	Israel	&	0.004	&	Angola	&	0.010	&	Singapore	\\
9	&	0.013	&	U Arab Em	&	0.015	&	C d'Ivoire	&	0.004	&	Spain	&	0.009	&	Germany	\\
10	&	0.012	&	Qatar	&	0.014	&	Japan	&	0.003	&	France	&	0.008	&	New Zealand	\\
\hline
\end{tabular}
\label{tab4}\label{table4}
\end{table}

\end{document}